\documentclass[%
reprint,
superscriptaddress,
 amsmath,amssymb,
 aps,
pra,
]{revtex4-2}

\usepackage{graphicx}
\usepackage{dcolumn}
\usepackage{bm}
\usepackage{hyperref}
\hypersetup{
    colorlinks=true,
    linkcolor=black,
    citecolor=blue,
    urlcolor=black
    }

\begin{document}

\preprint{APS/123-QED}

\title{Harnessing quantum light for microscopic biomechanical imaging of cells and tissues}

\author{Tian Li}
\affiliation{Department of Chemistry and Physics, University of Tennessee at Chattanooga, Chattanooga, TN 37403, USA}
\affiliation{UTC Research Institute, University of Tennessee at Chattanooga, Chattanooga, TN 37403, USA}

\author{Vsevolod Cheburkanov}
\affiliation{Department of Biomedical Engineering, Texas A\&M University, College Station, Texas 77843, USA}

\author{Vladislav V. Yakovlev}
\affiliation{Department of Biomedical Engineering, Texas A\&M University, College Station, Texas 77843, USA
}%
\affiliation{Institute for Quantum Science and Engineering, Texas A$\&$M University, College Station, TX 77843, USA
}%
\affiliation{Department of Physics and Astronomy, Texas A$\&$M University, College Station, TX 77843, USA
}%

\author{Girish S. Agarwal}
\affiliation{Institute for Quantum Science and Engineering, Texas A$\&$M University, College Station, TX 77843, USA
}%
\affiliation{Department of Physics and Astronomy, Texas A$\&$M University, College Station, TX 77843, USA
}%
\affiliation{Department of Biological and Agricultural Engineering, Texas A$\&$M University, College Station, TX 77843, USA
}%

\author{Marlan O. Scully}
\email{scully@tamu.edu}
\affiliation{Institute for Quantum Science and Engineering, Texas A$\&$M University, College Station, TX 77843, USA
}%
\affiliation{Department of Physics and Astronomy, Texas A$\&$M University, College Station, TX 77843, USA
}%
\affiliation{Department of Electrical and Computer Engineering, Princeton University, Princeton, NJ 08540, USA
}%


\begin{abstract}
The biomechanical properties of cells and tissues play an important role in our fundamental understanding of the structures and functions of biological systems at both the cellular and subcellular levels. Recently, Brillouin microscopy, which offers a label-free spectroscopic means of assessing viscoelastic properties in vivo, has emerged as a powerful way to interrogate those properties on a microscopic level in living tissues. However, susceptibility to photo-damage and photo-bleaching, particularly when high-intensity laser beams are used to induce Brillouin scattering, poses a significant challenge. This article introduces a transformative approach designed to mitigate photo-damage in biological and biomedical studies, enabling non-destructive, label-free assessments of mechanical properties in live biological samples. By leveraging quantum-light-enhanced stimulated Brillouin scattering (SBS) imaging contrast, the signal-to-noise ratio is significantly elevated, thereby increasing sample viability and extending interrogation times without compromising the integrity of living samples. The tangible impact of this novel methodology is evidenced by a notable three-fold increase in sample viability observed after subjecting the samples to three hours of continuous squeezed-light illumination, surpassing the traditional coherent light-based approaches. The quantum-enhanced SBS imaging holds promise across diverse fields, such as cancer biology and neuroscience where preserving sample vitality is of paramount significance. By mitigating concerns regarding photo-damage and photo-bleaching associated with high-intensity lasers, this technological breakthrough expands our horizons for exploring the mechanical properties of live biological systems, paving the way for a new era of research and clinical applications.
\end{abstract}

\maketitle

\section{Introduction}

Mechanical interactions regulate a wide range of fundamental biological activities, including morphogenesis~\cite{bosveld2012mechanical,behrndt2012forces}, cell migration~\cite{fischer2012stiffness,friedl2012new,yamada2019mechanisms}, polarization~\cite{dufort2011balancing,ridley2003cell}, proliferation~\cite{jaalouk2009mechanotransduction,helmlinger1997solid}, and cell fate~\cite{engler2006matrix,wang2009mechanotransduction}. Local viscoelastic properties of molecular, sub-cellular, and cellular structures play a crucial role in defining those forces and their outcomes~\cite{baker2009extracellular,chaudhuri2020effects,vining2017mechanical}. A prominent example is cancer: abnormal elastic properties of cancer cells and local extracellular matrix (ECM) serve as unique markers for cancer~\cite{jonietz2012mechanics,troyanova2019differentiating}, while local mechanical properties control cancer progression~\cite{lu2012extracellular} and metastasis~\cite{friedl2004dynamic,friedl2019rethinking,friedl2011cancer}. Many diseases involve pathological changes in tissue stiffness~\cite{bao2003cell,savin2011growth,blacher2005large}, providing an early diagnostic tool for disease development.

More broadly, fundamental understanding of embryonic development~\cite{oates2009quantitative}, mechanotransduction~\cite{gillespie2001molecular}, \textit{in situ} tissue regeneration~\cite{gaharwar2020engineered}, drug delivery~\cite{ghosh2007micromechanical}, infection diseases~\cite{roos2010physical,karampatzakis2017probing}, and advanced biomaterials~\cite{langer2004designing,koski2013non,balaban2001force} calls for the development of new tools and methodologies for assessment of viscoelastic properties on a microscopic (sub-cellular) scale~\cite{bao2003cell,greenleaf2003selected,discher2009biomechanics,ingber2006mechanical,mouw2014extracellular}. Microscopic biomechanics plays an important role in understanding structures and functions of biological systems at the cellular and subcellular level. Biomechanics of single cells, subcellular components, and biomolecules have vastly contributed to the development of biomedical sciences~\cite{bao2003cell,list2005deregulated,paszek2005tensional,radisky2005rac1b,suresh2007biomechanics,huwart2008magnetic,butcher2009tense,scarcelli2012brillouin,stachs2012spatially,scarcelli2013brillouin,comoglio2005cancer,paszek2005tensional}. For example, recent studies suggested that the stiffness of ECM might affect the behavior of tissue by modulating cell contractility, which is crucial in tumorigenesis~\cite{list2005deregulated,paszek2005tensional}. Similarly, substrate stiffness of neuron cells plays important role in neuronal development, growth, and health~\cite{spedden2013neuron}. There pathogenesis problems are accompanied with complex cellular level mechanochemical processes. For instance, the matrix stiffness (fibrosis of ECM) can activate both Rho protein signaling and Erk (extracellular signal-regulated kinases) signaling pathways via integrin clustering~\cite{comoglio2005cancer}. As a result, tumorigenic processes, including changes in cell contractility, depolarization, and proliferation, herald transformation of normal epithelial cells into malignant ones~\cite{paszek2005tensional}. Local viscosity is equally important since material transport and metabolic reactions in living cells are limited by diffusion. Therefore, gaining insights into microscopic mechanical interactions and local viscoelastic properties enhances our fundamental understanding of biological processes and establishes a foundation for advancements in disease diagnosis, prevention, and therapeutic strategies. Given that these dynamic processes occur on a microscopic scale in three dimensions, there exists a notable technological gap in instruments capable of non-invasive, label-free assessment of these properties, as previously recognized~\cite{bao2003cell,van2003biomechanics}.

\begin{figure*}[t]
     \centering
     \includegraphics[width=1.0\linewidth]{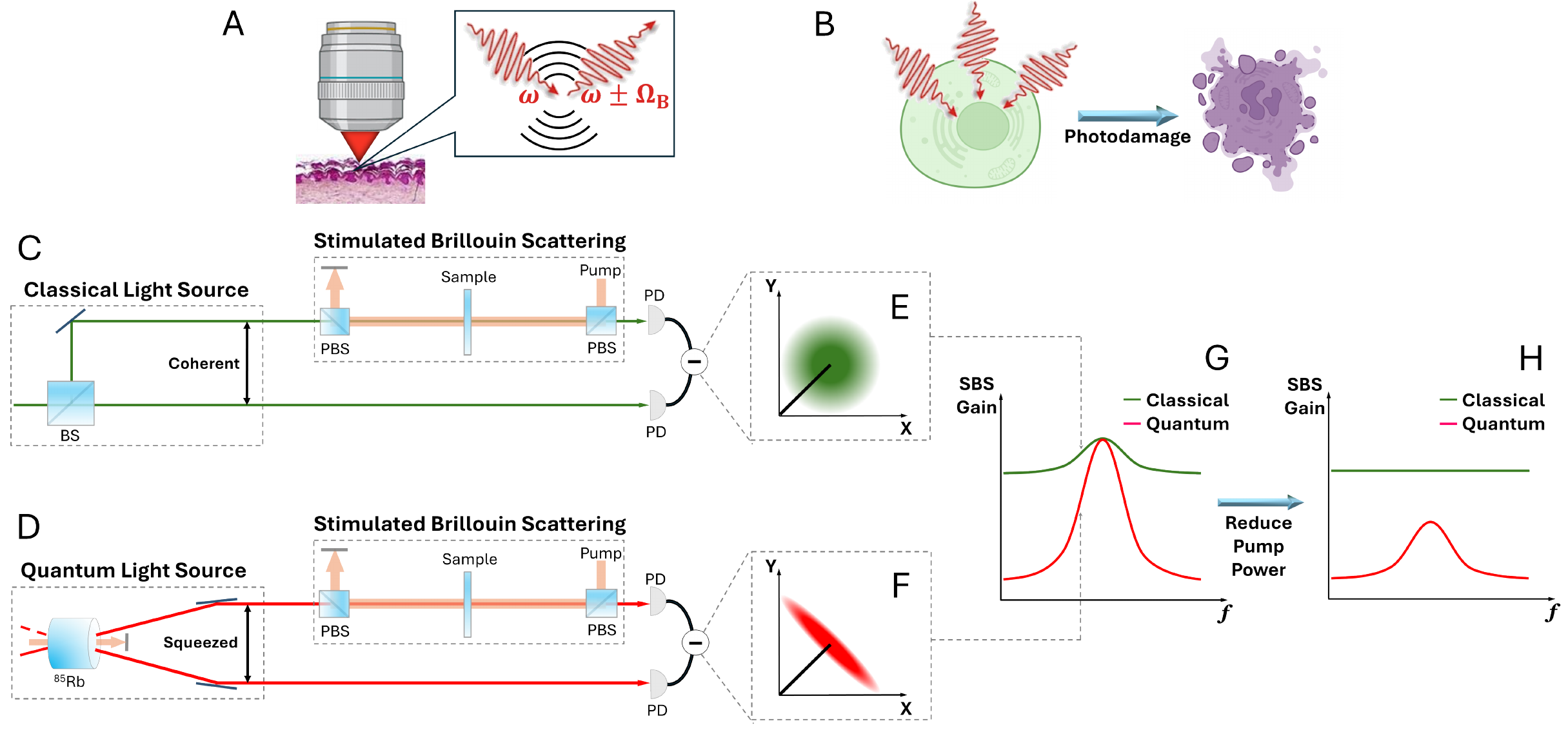}
    \caption{{\bf (A)} Schematic of conventional Brillouin microscopy, which originates from the inelastic optical scattering from acoustic phonons with characteristic resonance of $\Omega_\text{B}$ in a medium. {\bf (B)} Photodamage to a biological sample is induced by excessive input light intensity used to yield sufficient SNR. {\bf (C) \& (D)} Schematics of the stimulated Brillouin scattering (SBS) setups used in this work for classical light source in {\bf (C)} and quantum light source in {(D)}. {\bf (E) \& (F)} Phase space representations of the noise properties for the coherent light source in {\bf (E)} and squeezed light source in {\bf (F)}. {\bf (G) \& (H)} Schematics of the SNRs of SBS gain obtained using coherent (green curve) and squeezed (red curve) light sources with high pump power in {\bf (G)} and low pump power in {\bf (H)}. 
     \label{fig:Conceptual}}
\end{figure*}

Optical methods for assessing the elastic properties of cells and tissues are appealing due to their non-invasive nature and suitable spatial resolution. Brillouin spectroscopy, which originates from the inelastic optical scattering from acoustic phonons in a medium (as shown in Fig.~\ref{fig:Conceptual}A)), is among the oldest optical techniques being used for assessing mechanical properties of biological samples~\cite{randall1979brillouin}. Brillouin microscopy has been recognized as an exceptional technology for advancing molecular mechanobiology. Named the 2022 science story by \textit{The Guardian}~\cite{Guardian}, it is the only tool that provides a non-invasive assessment of local viscoelastic properties in 3-D with unprecedented spatial resolution~\cite{prevedel2019brillouin,poon2020brillouin,palombo2019brillouin,elsayad2019brillouin,singaraju2019brillouin}. While scanning tandem interferometers, pioneered by Sandercock~\cite{sandercock2005trends}, provided high-quality Brillouin spectra and were originally sought to measure viscoelastic properties of non-biological materials~\cite{dil1982brillouin}, assessing small variations of those properties in highly scattering biological materials on a microscopic scale has proven to be challenging. The key technical obstacles, which have been thoroughly discussed~\cite{scarcelli2015noncontact,meng2016seeing,kennedy2017emergence}, pertain to the sensitivity, accuracy, and speed of these measurements. These issues are fundamentally linked to the signal-to-noise ratio  (SNR), which is ultimately constrained by the shot-noise limit (SNL) defined by the power of the incident laser beam. Unfortunately, living cells cannot withstand high-power laser beams for extended period of time~\cite{hawkins2006role,denton2006damage}. Although this issue can be partially alleviated by using longer wavelength excitation, significant heating effects are inevitable~\cite{bixler2014assessment}, hence limiting the sensitivity, accuracy, and speed of Brillouin microscopy measurements.

To enhance acquisition speed and spatial resolution while reducing elastic scattering background, \textit{stimulated} Brillouin scattering (SBS) was proposed and first observed by Chiao \textit{et al.}~\cite{PhysRevLett.12.592}. Since then SBS has emerged as a powerful tool for investigating the mechanical properties of biological samples at the microscopic level~\cite{yang2023pulsed,merklein2022100,remer2020high,ballmann2017impulsive,ballmann2015stimulated}. One of the primary applications of SBS in biological samples is in the study of tissue bio-mechanics~\cite{palombo2019brillouin,bevilacqua2019imaging,raghunathan2017evaluating,antonacci2016biomechanics,palombo2014biomechanics}. These measurements provide valuable insights into the elastic properties of tissues, such as stiffness and viscosity, which are crucial indicators of tissue health and pathology~\cite{poon2020brillouin,scarcelli2015noncontact}. In the context of biomedical imaging, SBS-based techniques offer non-destructive and label-free means to assess the mechanical properties of biological structures~\cite{shi2023non,koski2005brillouin}. For instance, in ophthalmology, SBS has been employed to measure the stiffness of the cornea, aiding in the diagnosis and monitoring of diseases like keratoconus~\cite{shao2018effects}. Similarly, in cancer research, SBS has been utilized to characterize the mechanical properties of tumors, offering potential insights into tumor progression~\cite{margueritat2019high} and response to treatment~\cite{conrad2019mechanical}. Moreover, SBS techniques can be integrated into existing imaging modalities, such as optical coherence tomography~\cite{raghunathan2017evaluating}, to enable multi-modal imaging with enhanced contrast and sensitivity. These integrations allow for comprehensive assessments of biological samples, combining structural and mechanical information to elucidate complex biological processes~\cite{palombo2014mechanical}. Furthermore, the non-invasive nature of SBS-based techniques makes them particularly attractive for in vivo applications, facilitating real-time monitoring of tissue dynamics and responses to external stimuli~\cite{zhang2016line}. By probing the mechanical properties of living organisms at the cellular and sub-cellular levels, SBS holds promise for advancing our understanding of bio-mechanical phenomena in health and disease~\cite{troyanova2019differentiating}.

Despite the versatile nature of SBS, it is not without limitations, one of which is its susceptibility to photo-damage and photo-bleaching, especially when high-intensity laser beams are used to induce Brillouin scattering~\cite{yang2023pulsed,remer2020high}. The intense laser light required for SBS measurements can induce thermal and photo-chemical effects in biological samples, leading to potential damage or alteration of the tissue under investigation~\cite{palombo2019brillouin} (as shown in Fig.~\ref{fig:Conceptual}B). Minimizing these effects while maintaining sufficient SNR poses a significant technical challenge in SBS-based experiments, particularly in live-cell imaging or \textit {in vivo} applications where sample viability is critical~\cite{garmire2017perspectives}.  This is where quantum light can play a vital role in SBS. In this article we represent a novel experimental scheme capable of significantly reducing photo-damage to biological samples through squeezed-light-enhanced SBS image contrast (i.e., image SNR). This reduced photo-damage is exemplified by a drastically improved sample viability. After 3 hours of continuous interrogation, we observed a 3-fold viability improvement by placing our live sample under squeezed light illumination as opposed to placing the live sample under coherent light illumination.

\textcolor{black}{In addition to our squeezed-light-enhanced SBS imaging approach, it is also important to acknowledge other quantum-enhanced nonlinear techniques in biosample investigations. Notable contributions include Bowen \textit{et al.}'s demonstration of quantum-enhanced coherent Raman spectroscopy, achieving a 35~\% improvement in signal-to-noise ratio for imaging molecular bonds within cells compared to conventional Raman microscopy~\cite{casacio2021quantum}. Bowen and colleagues also used squeezed light with 75~\% reduced amplitude noise for microrheology measurements in yeast cells, surpassing the quantum noise limit by 42~\% while tracking lipid granules in real time~\cite{taylor2013biological}. A comprehensive review on quantum metrology and its applications in biology by Taylor and Bowen is available in Ref.~\cite{taylor2016quantum}. Additionally, Andersen \textit{et al.} utilized squeezed-light-enhanced stimulated Raman scattering to probe Raman shifts in polymer samples, achieving a quantum-enhanced signal-to-noise ratio (SNR) of 3.60~dB relative to the shot-noise limited SNR~\cite{de2020quantum}.
Another significant and highly relevant advancement in biosample investigations is the work on entangled two-photon absorption spectroscopy by Goodson and colleagues~\cite{varnavski2022quantum,schlawin2017entangled,burdick2021enhancing,varnavski2020two,villabona2020measurements,varnavski2017entangled}.  In particular, they demonstrated imaging capabilities at a low excitation intensity of $10^7$ photons/s, which is 6 orders of magnitude lower than the excitation level for the classical two-photon imaging~\cite{varnavski2020two}. This represents a major advancement given that two-photon absorption spectroscopy is extensively used for deep tissue imaging~\cite{schlawin2018entangled,dorfman2016nonlinear}. Furthermore, their recent publication~\cite{varnavski2023colors} offers valuable experimental and theoretical insights into the benefits of using entangled photons to obtain quantum spectra of complex molecules.}

Our approach employs a standard ``modulation-demodulation'' approach in conjunction with a ``balanced detection'' scheme, enabling the detection of weak signals within a spectral range where the noise level is constrained by shot noise, akin to the techniques used in recent demonstrations~\cite{li2022quantum,remer2020high,ballmann2015stimulated}. As depicted in Figs.~\ref{fig:Conceptual}C and~\ref{fig:Conceptual}D, during the process of SBS, counter propagating pump and probe beams with frequencies $\omega_1$ and $\omega_2$ and intensities $\text{I}_1$ and $\text{I}_2$ overlap in the sample, efficiently interacting with a longitudinal acoustic phonon of frequency $\Omega_{\text{B}}$. \textcolor{black}{Note that this counter-propagating pump-probe configuration is suitable for relatively thin biological samples that are transparent/non-absorptive at 795~nm with minimal scattering, so that quantum correlations in the squeezed light can be preserved after transmission.} When $\omega_2$ is scanned around the Stokes frequency ($\omega_1-\Omega_{\text{B}}$), the probe intensity $\text{I}_2$ exhibits a stimulated Brillouin gain $\text{I}_{\text{B}}\propto \text{I}_1 \text{I}_2$ due to wave resonance. At the same time the pump intensity $\text{I}_1$ at $\omega_1$ undergoes a stimulated Brillouin loss ($-\text{I}_{\text{B}}$). Hence conversely, if $\omega_2$ is scanned around the anti-Stokes frequency ($\omega_1 + \Omega_{\text{B}}$), the loss effect would be observed on the probe intensity $\text{I}_2$. The accompanying noise level, $\delta \text{I}_{\text{B}}$, of the stimulated Brillouin gain can be expressed as $\delta \text{I}_{\text{B}}\propto \beta \sqrt{\text{I}_{\text{2}}}$, where $\beta = 1$ for two coherent beams are used in the balanced detection as shown in Fig.~\ref{fig:Conceptual}C, and $\beta < 1$ for squeezed beams are used in the balanced detection as shown in Fig.~\ref{fig:Conceptual}D.
The phase space representations~\cite{scully1997quantum} of the noise properties for the coherent and squeezed light sources are illustrated in Figs.~\ref{fig:Conceptual}E and~\ref{fig:Conceptual}F, respectively. Due to the inherent strong quantum correlations between the squeezed beams, their amplitude/intensity difference noise level, represented by the red ellipse in Fig.~\ref{fig:Conceptual}F, is significantly reduced with respect to the shot noise level in the coherent case, represented by the green circle in Fig.~\ref{fig:Conceptual}E. When the pump beam intensity $\text{I}_1$ is sufficiently high, then both the classical and quantum cases would exhibit appreciable SNR, as depicted in Fig.~\ref{fig:Conceptual}G by the green and red curves, respectively. However, if one wishes to reduce photo-damage to the biological sample by lowering the pump beam intensity $\text{I}_1$ while maintaining the probe beam intensity $\text{I}_2$, a point will be reached where the Brillouin gain, $\text{I}_{\text{B}} \propto \text{I}_1 \text{I}_2$, becomes weaker than the shot noise level, $\delta \text{I}_{\text{B}}\propto \sqrt{\text{I}_{\text{2}}}$, denoted by the green curve in Fig.~\ref{fig:Conceptual}H, only the squeezed beams would give rise to an appreciable improvement in SNR, as indicated by the red curve in Fig.~\ref{fig:Conceptual}H. The primary measurements in this work are the stimulated Brillouin gain, $\text{I}_{\text{B}} \propto \text{I}_1 \text{I}_2$, and its associated noise, $\delta \text{I}_{\text{B}} \propto \beta \sqrt{\text{I}_2}$, where $\beta < 1$, which serve as our contrast for imaging biological samples.

Our quantum enhanced SBS scheme has potential implications for a wide range of fields, including biomedicine, tissue engineering, and regenerative medicine where sample viability is crucial~\cite{shi2023non}. It therefore will broaden our capabilities for investigating the mechanical properties of biological systems, opening new avenues for research and clinical applications.

\begin{figure*}[t]
    \centering
    \includegraphics[width=1.0\linewidth]{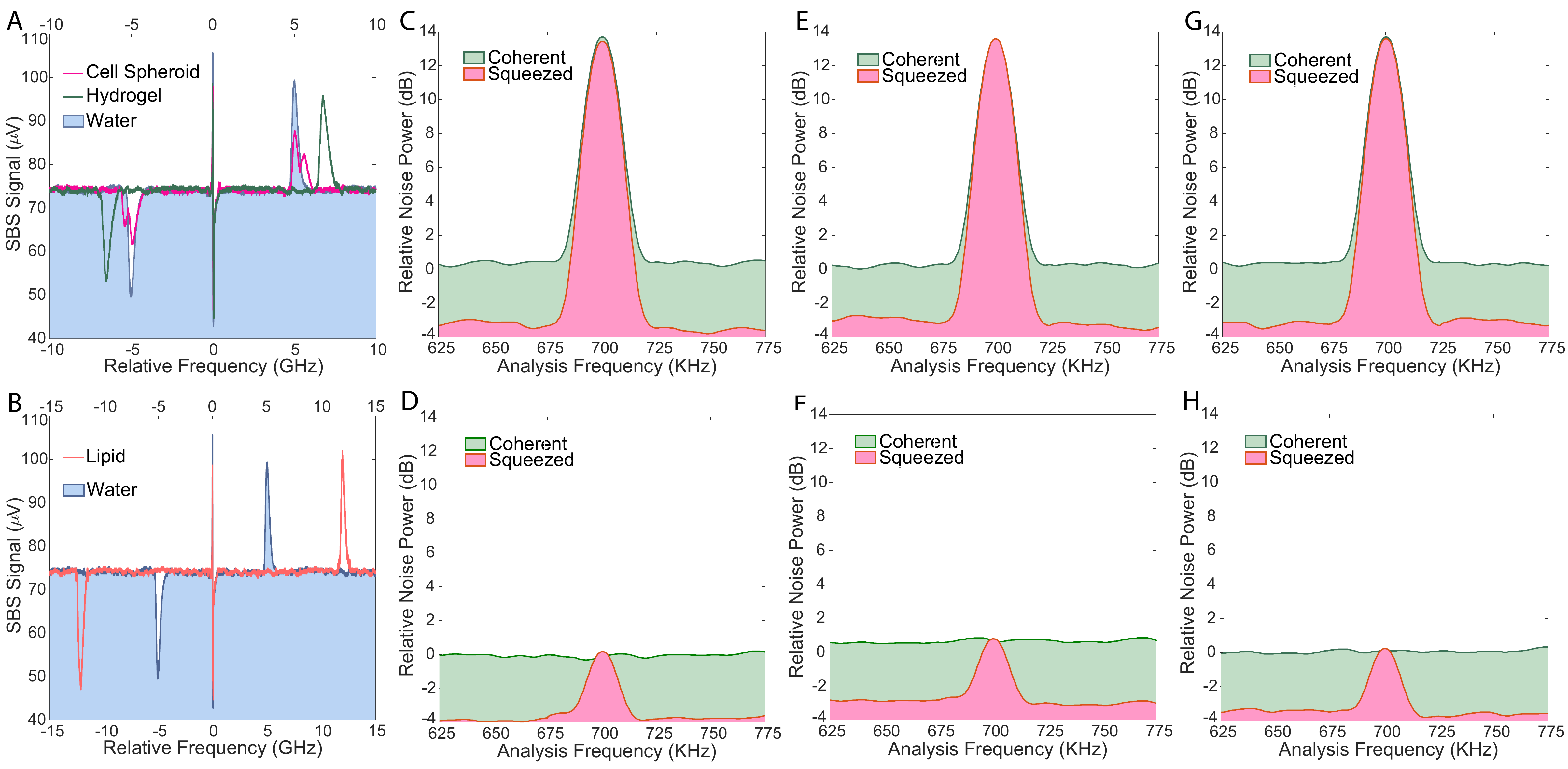}
    \caption{
    {\bf (A)} SBS spectra acquired by a lock-in amplifier for components in fixed 4T1 breast cancer cell spheroids suspended in hydrogel, where the blue, magenta, and green curves denote water, cell spheroid, and hydrogel spectra respectively. {\bf (B)} SBS spectra acquired by a lock-in amplifier for components in crushed drosophila brain, where the blue and red curves denote water and lipid spectra respectively. For both  {\bf (A) \& (B)}, the optical powers of the coherent probe and pump beams at the sample are 500~$\mu$W and 40~mW respectively. {\bf (C) \& (D)} SBS gains acquired by a spectrum analyzer when the SBS pump and probe beams are locked at the \textit{hydrogel} resonance for pump powers at {\bf (C)} 30~mW and {\bf (D)} 7~mW respectively. {\bf (E) \& (F)} SBS gains acquired by a spectrum analyzer when the SBS pump and probe beams are locked at the \textit{cell spheroid} resonance for pump powers at {\bf (E)} 55~mW and {\bf (F)} 15~mW respectively. {\bf (G) \& (H)} SBS gains acquired by a spectrum analyzer when the SBS pump and probe beams are locked at the \textit{lipid} resonance for pump powers at {\bf (G)} 40~mW and {\bf (H)} 12~mW respectively. For all the SBS gains shown in {\bf (C) -- (H)}, the SBS probe power is fixed at 700~$\mu$W for both the coherent probe and twin probe, which are denoted by the green and red curves respectively. Clear quantum noise reduction enabled SNR enhancement can be seen in all the graphs. 
    \label{fig:Spectra}}
\end{figure*}

\section{Results}

The quantum light source employed in the experiment is a two-mode intensity-difference squeezed light generated with the four-wave mixing (FWM) process in an atomic $^{85}$Rb vapor cell, which has proven to be a great platform for quantum sensing applications~\cite{li2022quantum,dorfman2021multidimensional,PhysRevApplied.15.044030,prajapati2021quantum,Dowran:18,Anderson:17,Pooser:15}. Major advantages of this FWM-based quantum light generation scheme include strong quantum correlations exhibited by greater than 6~dB two-mode quantum squeezing, and narrow-band ``twin beams'' generation with $\sim10$~MHz spectral line-width~\cite{PhysRevResearch.3.033095,Clark:2014vf,Glasser2012a}. This narrow line-width feature is extremely beneficial for the intended SBS experiment, where the spectral width of the light source must be well below the Brillouin resonance line-width, which is typically a few hundreds of MHz. The SNR enabled by the twin beams, with signal defined as the difference of photon numbers in the twin beams, is better than that for coherent beams by a factor of $\text{cosh}2r$, where $r$ is the well-known squeezing parameter used to characterize the two-mode squeezed state~\cite{NonlinearOptics}. This improvement in SNR consequently translates to quantum-enhanced image contrast as demonstrated in our proof-of-concept experiment on SBS spectroscopy~\cite{li2022quantum}.

In our prior work~\cite{li2022quantum}, two quantum-correlated ``twin beams'' of light, i.e., the ``probe'' and ``conjugate'' beams, are produced with the FWM process in an atomic $^{85}$Rb vapor cell. After the cell, the ``probe'' beam is overlapped with a counter-propagating laser beam (the ``pump'' beam for the SBS process labeled in Figs.~\ref{fig:Conceptual}C and~\ref{fig:Conceptual}D) at a sample holder to form a phase-matching geometry for the SBS process. The ``conjugate'' beam serves solely as a reference for the balanced detection scheme to cancel out common-mode noise in the twin beams. The quantum SBS spectroscopy setup (conceptualized in Fig.~\ref{fig:Conceptual}D) can be converted to a classical version (conceptualized in Fig.~\ref{fig:Conceptual}C) by replacing the probe and conjugate beams with two coherent beams having the same optical powers as the twin beams. In our experiment, the SBS signals (both gain and loss) are expected to appear at 700~KHz, \textcolor{black}{which is the sum frequency of the amplitude modulations on the two SBS beams~\cite{li2022quantum}}. The 700~KHz signal frequency is chosen since it is where the two-mode squeezing is expected to be the best~\cite{li2022quantum,PhysRevApplied.15.044030,doi:10.1063/5.0010909}. Both the SBS pump and probe lasers are locked to external cavities, and the relative frequency between them can be scanned \textcolor{black}{with 40~MHz spectral resolution so that the Brillouin shifts, i.e., the peak and dip of the Brillouin gain and loss, can be located.} \textcolor{black}{Also note that, our SBS signals are measured by a customized balanced detector, which subtracts away common-mode technical noise of the two input beams to better than 25~dB, so that noise level at 700~KHz (where the Brillouin signals appear) is shot-noise limited.} Other experimental details can be found in Methods.

We start with classically characterizing the SBS gain of different components in two biological samples. We use \textcolor{black}{the balanced detection scheme for this measurement for two reasons: 1). it enables the detection of weak signals within a spectral range where the noise level is constrained by shot noise, and 2). our quantum light is in a \textit{two-mode} squeezed state where the squeezing resides in the \textit{intensity-difference} of the two involving modes.} Figures~\ref{fig:Spectra}A and~\ref{fig:Spectra}B contain the SBS spectra of various components in two different samples -- 4T1 breast cancer cell spheroids in hydrogel in Fig.~\ref{fig:Spectra}A and drosophila brain tissue in Fig.~\ref{fig:Spectra}B, respectively. All spectra are obtained from a lock-in amplifier with 300~ms time constant, and the coherent probe beam is locked while the pump beam of the SBS process is scanned with 0.02~Hz scan frequency. From the water (distilled $\text{H}_2\text{O}$, T = 21~$^\circ$C) SBS spectra (denoted by the shaded blue areas in Figs.~\ref{fig:Spectra}A and~\ref{fig:Spectra}B) the Brillouin shift and the gain line-width are measured to be $\Omega_B/2\pi = 5.03 \textcolor{black}{\pm0.15}$~GHz and $\Gamma/2\pi = 287 \textcolor{black}{\pm23}$~MHz, which are in excellent agreement with previous experiments~\cite{li2022quantum,ballmann2015stimulated}. The dips on the left and peaks on the right of zero are the stimulated Brillouin loss and gain resonances respectively. The center features are caused by absorptive stimulated Rayleigh scattering. We repeated the same procedure and acquired the SBS spectra for cancer cell spheroid, hydrogel, and lipid, and denote them with magenta, green, and red curves respectively in Figs.~\ref{fig:Spectra}A and~\ref{fig:Spectra}B.

\begin{figure*}[t]
    \centering
    \includegraphics[width=1.0\linewidth]{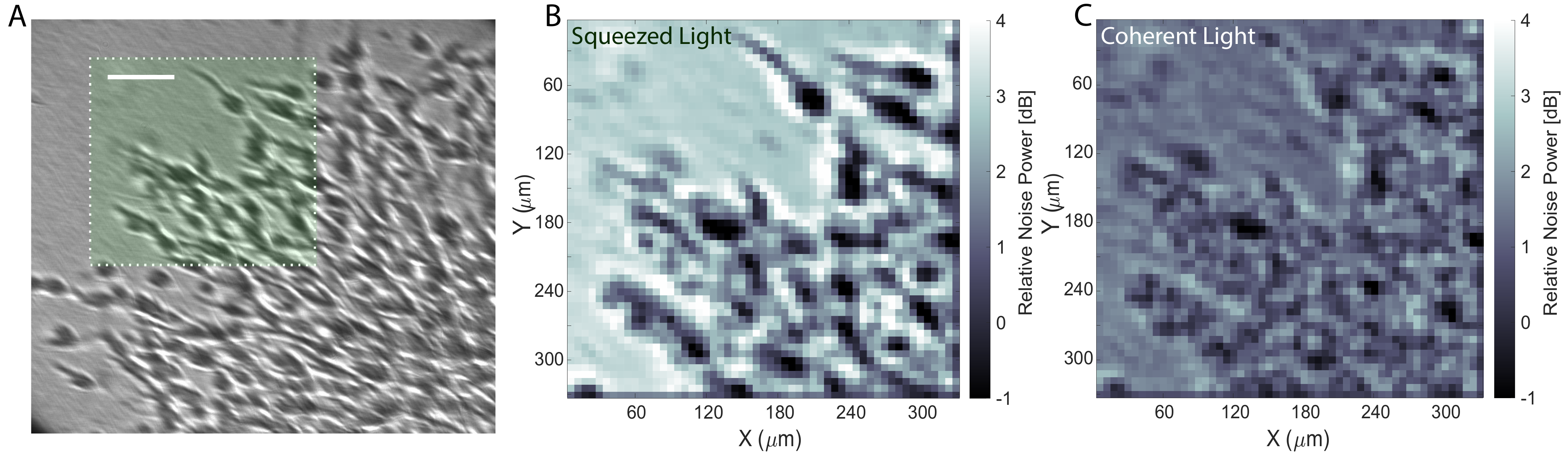}
    \caption{
     SBS images obtained with the two SBS lasers are locked at the gain peak of the $\textit{hydrogel}$ spectrum, , i.e., the peak at 6.7~GHz shown in Fig.~\ref{fig:Spectra}(A).
     {\bf (A)} 4T1 breast cancer cell spheroids under bright field illumination, the length of the white bar indicates 100~$\mu$m. {\bf (B)} Quantum-enhanced SBS images of the green squares shown in (B), obtained with the squeezed twin beams. {\bf (C)} Classical counterparts of (B), obtained with two coherent beams. The image resolution of $\sim5~\mu$m is determined by the focal spot size of the SBS beams, which is calculated using the numerical aperture $NA = 0.09$ of the focusing optics. Each image has $55\times55$ pixels with pixel size of 6~$\mu$m. The image contrast (i.e., SNR) is displayed by the relative noise power (in dB) indicated by the color bar. Images are sharpened using Matlab's `imsharpen' function with `Radius' = 2, `Amount' = 1, and `Threshold' = 0.1.
    \label{fig:Hydrogel}}
\end{figure*}

\begin{figure*}[t]
    \centering
    \includegraphics[width=1.0\linewidth]{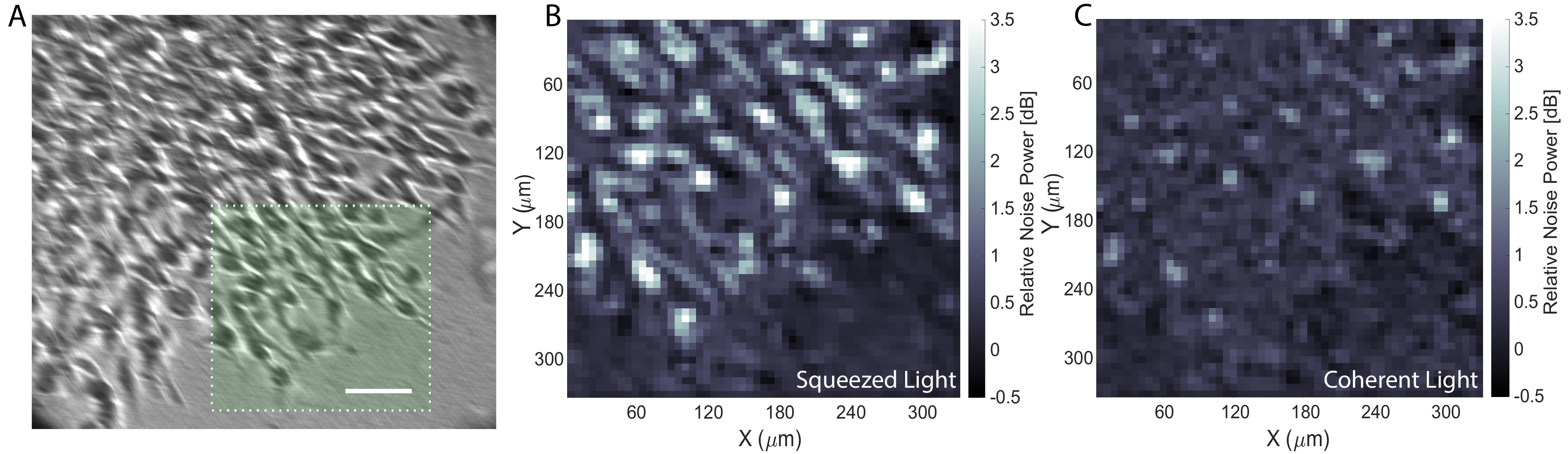}
    \caption{
    SBS images obtained with the two SBS lasers are locked at the side gain peak of the $\textit{cancer cell spheroid}$ spectrum, i.e., the peak at 5.6~GHz shown in Fig.~\ref{fig:Spectra}(A). {\bf (A)} 4T1 breast cancer cell spheroids under bright field illumination, the length of the white bar indicates 100~$\mu$m. {\bf (B)} Quantum-enhanced SBS images of the green squares shown in (A), obtained with the squeezed twin beams. {\bf (C)} Classical counterparts of (B), obtained with two coherent beams. The image resolution of $\sim5~\mu$m is determined by the focal spot size of the SBS beams, which is calculated using the numerical aperture $NA = 0.09$ of the focusing optics. Each image has $55\times55$ pixels with pixel size of 6~$\mu$m. The image contrast (i.e., SNR) is displayed by the relative noise power (in dB) indicated by the color bar. Images are sharpened using Matlab's `imsharpen' function with `Radius' = 2, `Amount' = 1, and `Threshold' = 0.1.
    \label{Cell}}
\end{figure*}

Having characterized the classical SBS process, in the following we demonstrate the quantum-enhanced SBS spectra. To clearly demonstrate quantum-improved performance beyond the classical approach, we conducted the experiment both with the probe beam in a coherent state (labeled as `Classical Probe' in Fig.~\ref{Setup}C) and in the two-mode squeezed state (labeled as `Quantum Probe' in Fig.~\ref{Setup}C). The experimental layouts can be easily swapped between the two configurations simply by replacing the twin beams with two coherent beams. Figures~\ref{fig:Spectra}C and~\ref{fig:Spectra}D represent our experimental results for the quantum-enhanced SBS spectra of hydrogel. In order to acquire the spectra, both lasers are locked so that their frequency difference matches the Brillouin shift of hydrogel, which in our case is 6.7~GHz (see the green spectrum in Fig.~\ref{fig:Spectra}A). The data presented in Figs.~\ref{fig:Spectra}C and~\ref{fig:Spectra}D and the following subfigures have all been measured by a RF spectrum analyzer with a resolution bandwidth of 10~KHz and a video bandwidth of 10~Hz. \textcolor{black}{With these bandwidths, the shot noise levels indicated by the green curves are around -67~dBm, whereas the electronic noise floor is around -81~dBm, and there is negligible contribution from the stray SBS pump beam to the detection noise}. We present the spectra for the Brillouin gain of hydrogel using coherent beams (green traces) and twin beams (red traces) with 700~$\mu$W probe power, while the SBS pump power is kept at 30~mW in Fig.~\ref{fig:Spectra}C and 7~mW in Fig.~\ref{fig:Spectra}D. It is clear from the spectra that the implementation of twin beams give rise to a significantly improved SNR ($\sim 3.5$~dB) of the SBS gain, and therefore an enhanced sensitivity of the Brillouin spectroscopy. We see in particular in Fig.~\ref{fig:Spectra}D that for a pump power of 7~mW, the Brillouin gain from two coherent beams is almost embedded in the shot-noise level, and only becomes pronounced when using twin beams. It is therefore clear that by using the two-mode squeezed light, it is possible to obtain Brillouin gain even for a continuous-wave (CW) pump laser power less than 7~mW. This is extremely beneficial for studying fragile biological samples where excessive optical power would damage the sample. Similar quantum-enhanced SNR can also be observed when the SBS laser beams are locked at the cancer cell spheroid gain peak in Figs.~\ref{fig:Spectra}E and~\ref{fig:Spectra}F for pump powers at 55~mW and 12~mW respectively; and when the SBS laser beams are locked at the lipid gain peak in Figs.~\ref{fig:Spectra}G and~\ref{fig:Spectra}H for pump powers at 30~mW and 7~mW respectively. For all the SBS gains shown in Figs.~\ref{fig:Spectra}C--H, the SBS probe power is fixed at 700~$\mu$W for both the classical probe and quantum probe.

\subsection{4T1 breast cancer cell spheroids in hydrogel}

We now demonstrate that our quantum-enhanced SBS spectroscopy can be utilized for microscopic imaging of biological samples. We use the SBS gain of various components of the sample to acquire a 2-Dimensional image of the whole sample. The first sample under investigation is fixed 4T1 breast cancer cell spheroids in hydrogel. The results are presented in Fig.~\ref{fig:Hydrogel}, where the Brillouin images were obtained with two SBS lasers were locked at the gain peak of the $\textit{hydrogel}$ spectrum, i.e., the peak at 6.7~GHz of the green curve shown in Fig.~\ref{fig:Spectra}A. Optical powers of the pump and probe beams are 7~mW and 700~$\mu$W respectively before the sample holder. Pixels in Fig.~\ref{fig:Hydrogel}B are registered with the probe beam being in the two-mode squeezed state, and pixels in Fig.~\ref{fig:Hydrogel}C are registered with the probe beam being in a coherent state. Obviously, the image contrast (i.e., the SNR) for the cell spheroids in Fig.~\ref{fig:Hydrogel}C is unappreciable due to the coherent light induced SBS gain of hydrogel is overwhelmed by the shot noise (see the green curve in Fig.~\ref{fig:Spectra}D). By using the two-mode squeezed light, however, more than 3.5~dB squeezed-light-enabled quantum advantage in image contrast can be clearly seen in Fig.~\ref{fig:Hydrogel}B (see the red curve in Fig.~\ref{fig:Spectra}D). In addition to locking the lasers at the hydrogel SBS gain peak, we also capture images by locking the SBS laser beams at the \textit{side} gain peak of the $\textit{cell spheroid}$ spectrum (whose \textit{main} peak overlaps with the water SBS gain peak), i.e., the \textit{secondary} peak at 5.6~GHz of the magenta curve in Fig.~\ref{fig:Spectra}A. The resulting images are shown in Fig.~\ref{Cell}. Optical powers of the pump and probe beams are now 15~mW and 700~$\mu$W respectively before the sample holder. Although the pump power is higher than that used in Fig.~\ref{fig:Hydrogel}, and the degradation of image quality is visible, more than 3~dB squeezed-light-enabled quantum advantage in image contrast can still be seen in Fig.~\ref{Cell}. The resolution of the images shown in Fig.~\ref{fig:Hydrogel} and Fig.~\ref{Cell} is $\sim 5~\mu$m, which is determined by the focal spot size of the SBS beams, calculated using the numerical aperture $NA = 0.09$ of the focusing optics, i.e., the two SBS beams both with $1/e^2 = 3$~mm diameter are focused at a same spot by two $f=16$~mm plano-convex lenses (see Fig.~\ref{Setup} in Method for a detailed optics layout).

It is worth pointing out that there are three gain peaks (two on the magenta curve, one on the green curves) on the spectrum of the fixed 4T1 breast cancer cell spheroids in hydrogel shown in Fig.~\ref{fig:Spectra}A, therefore in order to have a complete investigation, in addition to hydrogel and cell spheroids, we also used water SBS gain peak as contrast to form images. The results turned out to be subpar as the cell spheroids spectrum overlaps partially with the water spectrum.

\begin{figure*}[t]
    \centering
    \includegraphics[width=1.0\linewidth]{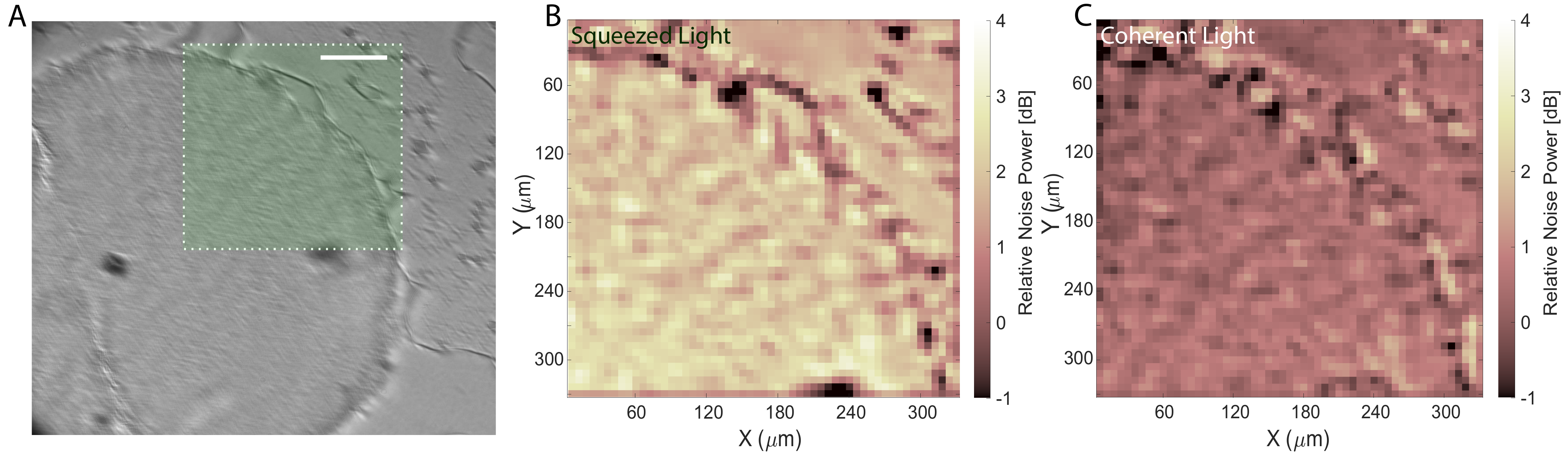}
    \caption{
     SBS images obtained with the two SBS lasers are locked at the gain peak of the $\textit{lipid}$ spectrum, i.e., the peak at 12~GHz shown in Fig.~\ref{fig:Spectra}(B). {\bf (A)} Drosophila brain tissue under bright field illumination, the length of the white bar indicates 100~$\mu$m. {\bf (B)} Quantum-enhanced SBS image of the green square shown in (A), obtained with the squeezed twin beams. {\bf (C)} Classical counterpart of (B), obtained with two coherent beams. The image resolution of $\sim5~\mu$m is determined by the focal spot size of the SBS beams, which is calculated using the numerical aperture $NA = 0.09$ of the focusing optics. Each image has $55\times55$ pixels with pixel size of 6~$\mu$m. The image contrast (i.e., SNR) is displayed by the relative noise power (in dB) indicated by the color bar. Images are sharpened using Matlab's `imsharpen' function with `Radius' = 2, `Amount' = 1, and `Threshold' = 0.1.
     \label{fig:Lipid}}
\end{figure*}

\subsection{Drosophila brain tissue}

We now use our quantum-enhanced SBS spectroscopy to image the second sample -- drosophila brain tissue. The SBS beams were focused on the central brain part of drosophila melanogaster central nervous system to achieve minimal optical track of $\sim 5~\mu$m. The SBS spectrum of the drosophila brain obtained from a lock-in amplifier is depicted in Fig.~\ref{fig:Spectra}B. The SBS gain and loss resonances at $\pm12$~GHz indicate that the sample is mainly composed of lipid as expected~\cite{antonacci2015quantification}. Using the gain peak of lipid as image contrast by locking the two lasers at $12$~GHz, and following the same operating procedure as previously described, the resulting SBS images are presented in Fig.~\ref{fig:Lipid}. We can clearly observe a close to 4~dB quantum enhanced image contrast in Fig.~\ref{fig:Lipid}B as opposed to its classical counterpart in Fig.~\ref{fig:Lipid}C. We attribute this improved quantum advantage, comparing to the $\sim3$~dB quantum advantage previously obtained with the sample of fixed 4T1 breast cancer cell spheroids, to the fact that the drosophila brain tissue is more transparent than the breast cancer cell spheroids for the SBS laser beams. Optical powers of the pump and probe beams here are 12~mW and 700~$\mu$W respectively before the sample holder.

\subsection{Improved sample viability under quantum light illumination}

\begin{figure*}[t]
    \centering
    \includegraphics[width=1.0\linewidth]{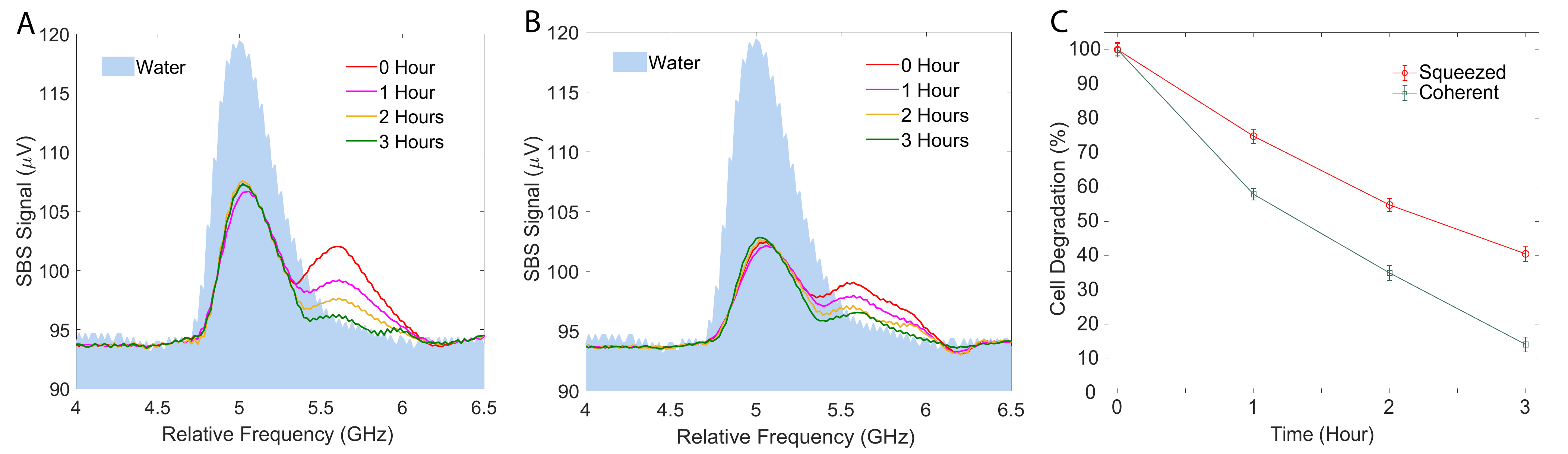}
    \caption{
    {\bf (A) \& (B)} Live 4T1 breast cancer cells degradation due to (A) classical coherent light and (B) quantum squeezed light illuminations for 3-hour interrogation time. {\bf (C)} Normalized side gain peak of the SBS spectra of the cancer cell spheroid for both the classical coherent light (green) and quantum squeezed light (red) illuminations.
    \label{Degradation}}
\end{figure*}

Provided that our two-mode squeezed light yields more than 3~dB quantum advantage over coherent light in image contrast, hence \textit{for a given SNR}, the pump power required in the quantum case would be less than that in the classical case by $\sim 50~\%$. This excitation power reduction would induce less photodamage, and thereby would significantly extend the interrogation time of sample. In order to prove that live samples can sustain longer under quantum light illumination, we compare the results obtained from \textit{live} 4T1 breast cancer cells interrogated by the two-mode squeezed light as well as by coherent light. The results are shown in Fig.~\ref{Degradation}, where Figs.~\ref{Degradation}A and~\ref{Degradation}B are for the probe beam (with a fixed optical power of 900~$\mu$W) being in a coherent state and in the two-mode squeezed state, respectively. Each curve in Figs.~\ref{Degradation}A and~\ref{Degradation}B is an average of 10 spectra acquired by a lock-in amplifier with 45~mW pump power in Fig.~\ref{Degradation}A and 24~mW pump power in Fig.~\ref{Degradation}B, so that \textit{the SNR of the cancer cell spheroid SBS gain are the same for these two cases}.

We placed our live cancer cells under consistent quantum and classical light illuminations for 3 hours, and measure their SBS spectra every one hour using a lock-in amplifier. We see from Fig.~\ref{Degradation}A that, the \textit{secondary} gain peak, which is the hallmark of the cancer cell spheroid, degrades much more aggressively than the ones shown in Fig.~\ref{Degradation}B during the 3-hour interrogation time. After 3 hours' coherent light exposure shown by Fig.~\ref{Degradation}A, the cancer cells under interrogation are almost completely dissolved into water, manifested by the overlapping of the green curve with the water SBS spectrum (shaded blue area). Whereas in the quantum case shown by Fig.~\ref{Degradation}B, there is still a clearly discernible secondary gain peak on the cancer cell SBS spectrum shown by the green curve after 3 hours' consistent squeezed light exposure. If we \textit{normalize these time-stamped secondary gain peaks with respect to their initial height} at the beginning of interrogation, which is plotted in Fig.~\ref{Degradation}C, we can clearly observe the difference -- the degradation rate of cancer cells' livelihood is much faster in the classical case than it is in the quantum case. After 3 hours of continuous interrogation, only 13~\% cancer cells survived under coherent light illumination, while 41~\% cancer cells survived under squeezed light illumination, thereby improving the cancer cell sample viability by 3 fold.

\section{Discussion and Outlook}

{\bf Uniqueness of our quantum light.}~The quantum advantage of our scheme is achieved by utilizing bright two-mode squeezed light with a spectral width in the range of 10~MHz, generated through the FWM process in atomic $^{85}$Rb vapor~\cite{PhysRevResearch.3.033095,Clark:2014vf,Glasser2012a}. It is this unique narrow-band feature of our two-mode squeezed light source that enables our quantum-enhanced SBS spectroscopy and imaging demonstration. In order for the SBS process to occur, the spectral width of the light source must be significantly below the Brillouin line-width of the components under investigation, and in this work, they are in the range of a few hundred MHz. Although the typical line-width of single-mode squeezed light generated from optical parametric oscillators (OPO) is usually in the range of a few MHz to tens of MHz, their photon flux typically ranges from $10^6$ to $10^9$ photons per second~\cite{park2024single}. This is several orders of magnitude lower than the photon flux of $10^{14}$ to $10^{16}$ photons per second that our two-mode squeezed light can achieve. These unique features of our two-mode squeezed light, generated through the FWM process in atomic $^{85}$Rb vapor, make it an exceptional quantum light source for SBS spectroscopy and imaging.

{\bf Image acquisition time.}~The quantum-enhanced SBS spectra shown in Fig.~\ref{fig:Spectra} were obtained from a spectrum analyzer with a 10~KHz resolution bandwidth, 10~Hz video bandwidth, and a 1-second sweep time to scan a frequency span of 150~KHz (from 625~KHz to 775~KHz). Notably, the sweep time can be significantly accelerated by using the ``zero span'' mode of the spectrum analyzer, combined with either reducing the resolution bandwidth or increasing the video bandwidth. In our case, the sweep time can be reduced to 2.0~ms while operating in the ``zero span'' mode with a 3~KHz resolution bandwidth and 300~Hz video bandwidth. To demonstrate the viability of our scheme with a much faster sweep time, we retook the SBS image of the 4T1 breast cancer cells in hydrogel using the ``zero span'' mode. The resulting image was essentially the same as the one shown in Fig.~\ref{fig:Hydrogel}A, albeit noisier due to the video bandwidth being 30 times wider. This implies that our acquisition rate is not fundamentally limited by our scheme but is rather technically constrained by the instrument. This technical limitation can be readily overcome with the use of a more advanced RF spectrum analyzer, such as a real-time spectrum analyzer with a much faster data writing and read-out rate, or with a large memory that allows data to be processed locally. Consequently, the acquisition time would then be limited solely by the sweep time of the spectrum analyzer. 


{\bf Image spatial resolution.}~Regarding spatial resolution, note that we opted to use $f=16$~mm plano-convex lenses to focus the probe and pump beams of the SBS process on the sample, primarily due to concerns about loss and alignment. While objectives could focus the beams much more tightly, they would also introduce significant losses \textcolor{black}{due to reflection and limited aperture which are normally specified by manufacturers. Those losses can be of the order of 30 -- 50~\% even for high quality objectives,} which would inevitably degrade the quantum advantage. To improve the special resolution using our current setup, a practical approach is to use molded aspheric lenses to focus the two SBS beams on the sample. For instance, using an aspheric lens with an effective focal length of $f=4.5$~mm would yield a $1/e^2$ focal spot size of $1.5~\mu$m in diameter, assuming our $1/e^2$ beam diameter at the lens is 3~mm. This change would enhance our resolution by a factor of three (reduced from $5~\mu$m to $1.5~\mu$m). However, it is also important to understand the challenges associated with this spatial resolution improvement -- the tighter focus necessitates a more precise alignment, as the pump and probe beams of the SBS process must overlap within an excitation volume of merely $1.5~\mu$m in diameter and $4.5~\mu$m in length. Achieving this overlap requires meticulous adjustment and fine-tuning of the beam paths.

\textcolor{black}{{\bf Motion of live cells.}~The motion of cells and cellular compartment poses a significant challenge for many live-cell biological imaging applications. Spontaneous Brillouin microscopic imaging is often too slow for such imaging exactly for the reason of cellular motion. Stimulated Brillouin microscopy partially addresses this issue by increasing the acquisition speed of such imaging by several orders of magnitude, but at the same time introduces the risk of cellular photodamage. Therefore, achieving stimulated Brillouin microscopy with reduced photodamage is crucial, and this can only be realized through quantum light spectroscopy. For the sake of comparison, we focus in this work on imaging biological samples where cellular motion is not significant, which allows us to capture the distribution of mechanical properties throughout the entire acquisition time.}

\section*{Methods}



In this work, we adopted the same methodology employed in our prior proof-of-principle demonstration~\cite{li2022quantum}, which we elaborate as follows.

\begin{figure*}[]
    \begin{center}
    \includegraphics[width=1\linewidth]{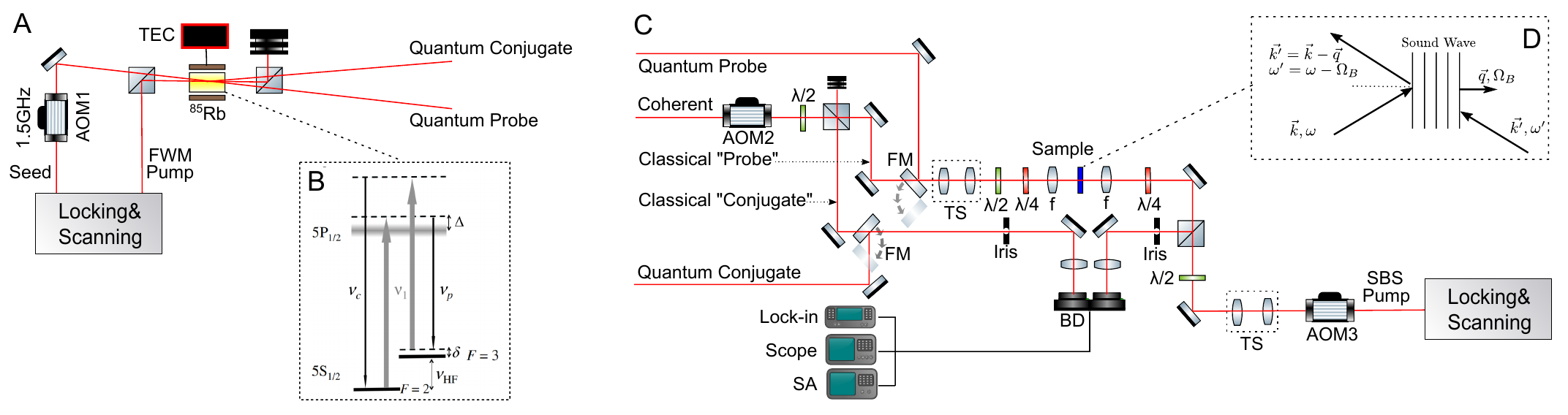}
    \caption{\textcolor{black}{
     {\bf (A)} Experimental setup for the bright two-mode squeezed light generation based on FWM in $^{85}$Rb atomic vapor. See text for a detailed description.  AOM: acousto-optic modulator, TEC: thermo-electric coupler. {\bf (B)} Level structure of the D1 transition of $^{85}$Rb atom. The optical transitions are arranged in a double~-~$\Lambda$ configuration, where $\nu_p$, $\nu_c$ and $\nu_1$ stand for probe, conjugate and pump frequencies, respectively, fulfilling $\nu_p$ +  $\nu_c$ =  $2\nu_1$ and $\nu_c - \nu_p = 2\nu_{HF}$. The width of the excited state in the level diagram represents the Doppler broadened line. $\Delta$ is the one-photon detuning. $\nu_{\text{HF}}$ is the hyperfine splitting in the electronic ground state of $^{85}$Rb. {\bf (C)} Experimental setup for the SBS spectroscopy for both the quantum and classical configurations. FM: flip mirror, TS: telescope, BD: balanced detector, SA: RF spectrum analyser. {\bf (D)} Phase-matching diagram for the SBS process~\cite{NonlinearOptics}. The wave-vectors and frequencies for the pump, probe and sound wave are denoted by ($\vec{k}$, $\omega$), ($\vec{k^{\prime}}$, $\omega^{\prime}$) and ($\vec{q}$, $\Omega_B$) respectively.}
        \label{Setup}}
    \end{center}
\end{figure*}

{\bf Full experimental layout.}~The atomic $^{85}$Rb vapor is pumped by a strong ($\sim 500$~mW) narrow-band continuous-wave (CW) laser composed of an external cavity diode laser (ECDL) and a tapered amplifier (TA) (shown in Fig.~\ref{Setup}(A) as ``FWM Pump") at frequency $\nu_1$ ($\lambda = 795$~nm) with a typical line-width $\Delta \nu_1 < 1$~MHz. Applying an additional weak (in the range of a few hundreds $\mu$W) coherent beam (shown in Fig.~\ref{Setup}(A) as ``Seed") 
at frequency $\nu_p = \nu_1 - (\nu_{HF}+\delta)$, where $\nu_{HF}=3.036$~GHz and $\delta$ are the hyperfine splitting in the electronic ground state of $^{85}$Rb and the two-photon detuning ($\delta=5$~MHz in this work) respectively in Fig.~\ref{Setup}(B). The frequency difference between ``FWM Pump'' and ``Seed'' is acquired by double-passing an 1.5~GHz acousto-optic modulator (AOM) (shown in Fig.~\ref{Setup}(A) as ``AOM1"). Two pump photons are converted into a pair of twin photons, namely `probe $\nu_p$' and `conjugate $\nu_c$' photons, adhering to the energy conservation $2 \nu_1 = \nu_p + \nu_c$ (see the level structure in Fig.~\ref{Setup}(B)). The resulting twin beams are strongly quantum-correlated and are also referred to as \textit{bright two-mode squeezed light}~\cite{PhysRevA.78.043816}. The twin beams exhibit a intensity-difference squeezing of 7~dB measured by a balanced detector with customized photodiodes having 94~\% quantum efficiency at 795~nm, which is indicative of strong quantum correlations~\cite{PhysRevA.78.043816}.

The ``FWM Pump'' and ``Seed'' beams are combined in a polarizing beam splitter (PBS) and directed at an angle of $\sim0.3^\circ$ to each other into a 12.5~mm long vapor cell filled with isotopically pure $^{85}$Rb. The two beams are collimated with 700~$\mu$m and 400~$\mu$m $1/e^2$ waists at the cell center respectively. The cell, with no magnetic shielding, is kept at $105^\circ$C by a thermo-electric coupler (TEC) and a PID feedback loop. The windows of the cell are anti-reflection coated on both faces, resulting in a transmission for the ``Seed'' beam of $\sim98$\% per window. After the $^{85}$Rb vapor cell, the ``FWM Pump'' and the twin beams (shown in Fig.~\ref{Setup}(A) as ``Quantum Probe" and ``Quantum Conjugate") are separated by a Glan-Laser polarizer, with $\sim$~$2\times10^5$~$:1$ extinction ratio for the Pump1. The ``Quantum Probe'' beam then passes through a telescope (TS) with an enlarged beam waist ($\sim 3$~mm) before focused (down to a $1/e^2$ beam waist of $\sim 5$~$\mu$m by a plano-convex lens with focal length $f= 16$~mm) and overlapped with a counter propagating laser beam (shown in Fig.~\ref{Setup}(C) as ``SBS Pump", a same type of ECDL as the ``FWM Pump", and having a $1/e^2$ beam waist of $\sim 6$~$\mu$m) at a homemade sample holder, to form a phase-matching geometry for the SBS process in the sample depicted in Fig.~\ref{Setup}(D). The sample holder consists of two glass microscope slides separated by 1~mm. Both $\lambda/2$ and $\lambda/4$ wave-plates are added in the probe beam path in order for the probe beam to be reflected as much as possible by the PBS into one port of the balanced detector (BD). Therefore in this configuration, the probe beam is linearly polarized while the pump beam for the SBS process (the ``SBS Pump" in Fig.~\ref{Setup}(C)) is circularly polarized. The conjugate beam serves as a reference, and two flip mirrors (FM) are used for the introduction of two coherent beams so that the whole setup can be converted into a classical version. The pumps and probe beams are amplitude-modulated by three AOMs at 300~KHz (AOM1) and 400~KHz (AOM2\&3) respectively. The SBS signal therefore is expected to appear at 700~KHz where the two-mode squeezing is expected to be the best~\cite{PhysRevApplied.15.044030,doi:10.1063/5.0010909}. The balanced detection uses two coherent beams and a balanced detector, which subtracts away common-mode noise to better than 25~dB, therefore contributions from low-frequency technical noise can be eliminated, so that the noise level at the modulation frequency (where the signal occurs) is shot-noise limited. There is no contribution from the stray pump light to the detection noise as there is a sufficient angle separation ($\sim0.3^\circ$) between the twin beams and the ``FWM Pump'' beam, and there are multiple irises in the paths of the twin beams to filter out the stray pump light.

{\bf Brillouin frequency shift locking.}~In addition to the components shown in Fig.~\ref{Setup}(A), there are also frequency-locking optics and electronics for the probe and pump beams of the SBS process so that they can be locked and separated by the phonon frequency (i.e., ``Brillouin frequency shift'') of different biological components, which is in the range of tens of GHz. In this work, we use fringes from a room temperature Fabry-Perot cavity as the locking error signal and absorption lines from a room temperature natural abundant Rb cell as the locking reference for each laser beam. We change the frequency difference between the two beams by fixing the probe frequency (blue-tuned by a ‘one-photon detuning $\Delta$’ of 1.1~GHz with respect to the $^{85}$Rb $5\text{S}_{1/2}, \text{F} = 2 \rightarrow 5\text{P}_{1/2}$, D1 transition shown in Fig.~\ref{Setup}(B)) so that the FWM process can yield the best two-mode intensity-difference squeezing, while scanning the locking frequency of the ``SBS Pump'' with a minimal step of 40~MHz determined by the resolution of the scanning voltage provided by a DC power supply.

{\bf Two-beam modulation.}~Note that in principle only one modulation on the pump beam would be sufficient for attaining the SBS signal. However, one must be extremely careful to eradicate any contribution from the pump light to the detected signal, as any amount of residual pump leakage into the detector would appear as spurious SBS signal. To do this, Ref.~\cite{remer2020high} used a Rubidium-85 notch filter at the pump frequency. This approach, however, would not be practical in our scheme as our two-mode squeezed twin beams, i.e., the ``Quantum Probe'' and ``Quantum Conjugate'' beams, are only a few GHz separated from the ``FWM Pump'' frequency, therefore the use of any notch filter at the pump frequency would inevitably induce undesired atomic absorption at the probe and conjugate frequencies as well. This would significantly deteriorate the quantum correlations between the probe and conjugate beams, and eventually wear out the quantum advantage. The 2-beam modulation (pump and probe at 300~KHz and 400~KHz respectively) approach adopted in our scheme solved this issue, as the SBS signal appeared at the sum frequency 700~KHz, hence even if there is residual pump leakage into the detector, the ``spurious signal'' would only appear at 300~KHz.

{\bf Two-beam balanced detection.}~It is crucial to note that in our scheme, we used a ``two-mode" squeezed state, where squeezing resides in the ``intensity-difference" between the two involving modes. As opposed to the experimental complexity of a single-mode squeezed scheme where a homodyne measurement is needed to characterize the squeezing, and a phase-locking mechanism is needed to track the squeezed quadrature, our scheme only requires a balance detector so that an intensity-difference measurement can be obtained. Therefore a ``balanced coherent detection" where two coherent beams are used would be the appropriate classical counterpart to the quantum configuration in our scheme.

{\bf Microscopic image acquisition.}~To acquire the microscopic images in the main text, we use two translational stages with differential micrometer screws to automatically move the sample holder's position with a spatial scan step size of 6~$\mu$m in both directions. The images are obtained by scanning each pixel under the experimental conditions shown in Figs.~\ref{fig:Spectra}(D), \ref{fig:Spectra}(F), and \ref{fig:Spectra}(H).\\

\section*{acknowledgments}
We would like to thank the Robert A. Welch Foundation (A-1261 and A-1943), the Air Force Office of Scientific Research (FA9550-20-1-0366), the National Science Foundation (PHY-2013771), and the U.S. Department of Energy (DE-SC-0023103 and DE-AC36-08GO28308). V.V.Y. received partial funding from the Air Force Office of Scientific Research (FA9550-20-1-0366, FA9550-20-1-0367, FA9550-23-1-0599), the National Institutes of Health (NIH) (R01GM127696, R01GM152633, R21GM142107, and 1R21CA269099). This material is also based upon work supported by the NASA, BARDA, NIH, and USFDA, under Contract/Agreement No. 80ARC023CA002.






\bibliography{MyLibrary}

\begin{thebibliography}{110}%
\makeatletter
\providecommand \@ifxundefined [1]{%
 \@ifx{#1\undefined}
}%
\providecommand \@ifnum [1]{%
 \ifnum #1\expandafter \@firstoftwo
 \else \expandafter \@secondoftwo
 \fi
}%
\providecommand \@ifx [1]{%
 \ifx #1\expandafter \@firstoftwo
 \else \expandafter \@secondoftwo
 \fi
}%
\providecommand \natexlab [1]{#1}%
\providecommand \enquote  [1]{``#1''}%
\providecommand \bibnamefont  [1]{#1}%
\providecommand \bibfnamefont [1]{#1}%
\providecommand \citenamefont [1]{#1}%
\providecommand \href@noop [0]{\@secondoftwo}%
\providecommand \href [0]{\begingroup \@sanitize@url \@href}%
\providecommand \@href[1]{\@@startlink{#1}\@@href}%
\providecommand \@@href[1]{\endgroup#1\@@endlink}%
\providecommand \@sanitize@url [0]{\catcode `\\12\catcode `\$12\catcode `\&12\catcode `\#12\catcode `\^12\catcode `\_12\catcode `\%12\relax}%
\providecommand \@@startlink[1]{}%
\providecommand \@@endlink[0]{}%
\providecommand \url  [0]{\begingroup\@sanitize@url \@url }%
\providecommand \@url [1]{\endgroup\@href {#1}{\urlprefix }}%
\providecommand \urlprefix  [0]{URL }%
\providecommand \Eprint [0]{\href }%
\providecommand \doibase [0]{https://doi.org/}%
\providecommand \selectlanguage [0]{\@gobble}%
\providecommand \bibinfo  [0]{\@secondoftwo}%
\providecommand \bibfield  [0]{\@secondoftwo}%
\providecommand \translation [1]{[#1]}%
\providecommand \BibitemOpen [0]{}%
\providecommand \bibitemStop [0]{}%
\providecommand \bibitemNoStop [0]{.\EOS\space}%
\providecommand \EOS [0]{\spacefactor3000\relax}%
\providecommand \BibitemShut  [1]{\csname bibitem#1\endcsname}%
\let\auto@bib@innerbib\@empty
\bibitem [{\citenamefont {Bosveld}\ \emph {et~al.}(2012)\citenamefont {Bosveld}, \citenamefont {Bonnet}, \citenamefont {Guirao}, \citenamefont {Tlili}, \citenamefont {Wang}, \citenamefont {Petitalot}, \citenamefont {Marchand}, \citenamefont {Bardet}, \citenamefont {Marcq}, \citenamefont {Graner} \emph {et~al.}}]{bosveld2012mechanical}%
  \BibitemOpen
  \bibfield  {author} {\bibinfo {author} {\bibfnamefont {F.}~\bibnamefont {Bosveld}}, \bibinfo {author} {\bibfnamefont {I.}~\bibnamefont {Bonnet}}, \bibinfo {author} {\bibfnamefont {B.}~\bibnamefont {Guirao}}, \bibinfo {author} {\bibfnamefont {S.}~\bibnamefont {Tlili}}, \bibinfo {author} {\bibfnamefont {Z.}~\bibnamefont {Wang}}, \bibinfo {author} {\bibfnamefont {A.}~\bibnamefont {Petitalot}}, \bibinfo {author} {\bibfnamefont {R.}~\bibnamefont {Marchand}}, \bibinfo {author} {\bibfnamefont {P.-L.}\ \bibnamefont {Bardet}}, \bibinfo {author} {\bibfnamefont {P.}~\bibnamefont {Marcq}}, \bibinfo {author} {\bibfnamefont {F.}~\bibnamefont {Graner}}, \emph {et~al.},\ }\bibfield  {title} {\bibinfo {title} {Mechanical control of morphogenesis by fat/dachsous/four-jointed planar cell polarity pathway},\ }\href@noop {} {\bibfield  {journal} {\bibinfo  {journal} {Science}\ }\textbf {\bibinfo {volume} {336}},\ \bibinfo {pages} {724} (\bibinfo {year} {2012})}\BibitemShut {NoStop}%
\bibitem [{\citenamefont {Behrndt}\ \emph {et~al.}(2012)\citenamefont {Behrndt}, \citenamefont {Salbreux}, \citenamefont {Campinho}, \citenamefont {Hauschild}, \citenamefont {Oswald}, \citenamefont {Roensch}, \citenamefont {Grill},\ and\ \citenamefont {Heisenberg}}]{behrndt2012forces}%
  \BibitemOpen
  \bibfield  {author} {\bibinfo {author} {\bibfnamefont {M.}~\bibnamefont {Behrndt}}, \bibinfo {author} {\bibfnamefont {G.}~\bibnamefont {Salbreux}}, \bibinfo {author} {\bibfnamefont {P.}~\bibnamefont {Campinho}}, \bibinfo {author} {\bibfnamefont {R.}~\bibnamefont {Hauschild}}, \bibinfo {author} {\bibfnamefont {F.}~\bibnamefont {Oswald}}, \bibinfo {author} {\bibfnamefont {J.}~\bibnamefont {Roensch}}, \bibinfo {author} {\bibfnamefont {S.~W.}\ \bibnamefont {Grill}},\ and\ \bibinfo {author} {\bibfnamefont {C.-P.}\ \bibnamefont {Heisenberg}},\ }\bibfield  {title} {\bibinfo {title} {Forces driving epithelial spreading in zebrafish gastrulation},\ }\href@noop {} {\bibfield  {journal} {\bibinfo  {journal} {Science}\ }\textbf {\bibinfo {volume} {338}},\ \bibinfo {pages} {257} (\bibinfo {year} {2012})}\BibitemShut {NoStop}%
\bibitem [{\citenamefont {Fischer}\ \emph {et~al.}(2012)\citenamefont {Fischer}, \citenamefont {Myers}, \citenamefont {Gardel},\ and\ \citenamefont {Waterman}}]{fischer2012stiffness}%
  \BibitemOpen
  \bibfield  {author} {\bibinfo {author} {\bibfnamefont {R.~S.}\ \bibnamefont {Fischer}}, \bibinfo {author} {\bibfnamefont {K.~A.}\ \bibnamefont {Myers}}, \bibinfo {author} {\bibfnamefont {M.~L.}\ \bibnamefont {Gardel}},\ and\ \bibinfo {author} {\bibfnamefont {C.~M.}\ \bibnamefont {Waterman}},\ }\bibfield  {title} {\bibinfo {title} {Stiffness-controlled three-dimensional extracellular matrices for high-resolution imaging of cell behavior},\ }\href@noop {} {\bibfield  {journal} {\bibinfo  {journal} {Nature Protocols}\ }\textbf {\bibinfo {volume} {7}},\ \bibinfo {pages} {2056} (\bibinfo {year} {2012})}\BibitemShut {NoStop}%
\bibitem [{\citenamefont {Friedl}\ \emph {et~al.}(2012)\citenamefont {Friedl}, \citenamefont {Sahai}, \citenamefont {Weiss},\ and\ \citenamefont {Yamada}}]{friedl2012new}%
  \BibitemOpen
  \bibfield  {author} {\bibinfo {author} {\bibfnamefont {P.}~\bibnamefont {Friedl}}, \bibinfo {author} {\bibfnamefont {E.}~\bibnamefont {Sahai}}, \bibinfo {author} {\bibfnamefont {S.}~\bibnamefont {Weiss}},\ and\ \bibinfo {author} {\bibfnamefont {K.~M.}\ \bibnamefont {Yamada}},\ }\bibfield  {title} {\bibinfo {title} {New dimensions in cell migration},\ }\href@noop {} {\bibfield  {journal} {\bibinfo  {journal} {Nature Reviews Molecular Cell Biology}\ }\textbf {\bibinfo {volume} {13}},\ \bibinfo {pages} {743} (\bibinfo {year} {2012})}\BibitemShut {NoStop}%
\bibitem [{\citenamefont {Yamada}\ and\ \citenamefont {Sixt}(2019)}]{yamada2019mechanisms}%
  \BibitemOpen
  \bibfield  {author} {\bibinfo {author} {\bibfnamefont {K.~M.}\ \bibnamefont {Yamada}}\ and\ \bibinfo {author} {\bibfnamefont {M.}~\bibnamefont {Sixt}},\ }\bibfield  {title} {\bibinfo {title} {Mechanisms of 3d cell migration},\ }\href@noop {} {\bibfield  {journal} {\bibinfo  {journal} {Nature Reviews Molecular Cell Biology}\ }\textbf {\bibinfo {volume} {20}},\ \bibinfo {pages} {738} (\bibinfo {year} {2019})}\BibitemShut {NoStop}%
\bibitem [{\citenamefont {DuFort}\ \emph {et~al.}(2011)\citenamefont {DuFort}, \citenamefont {Paszek},\ and\ \citenamefont {Weaver}}]{dufort2011balancing}%
  \BibitemOpen
  \bibfield  {author} {\bibinfo {author} {\bibfnamefont {C.~C.}\ \bibnamefont {DuFort}}, \bibinfo {author} {\bibfnamefont {M.~J.}\ \bibnamefont {Paszek}},\ and\ \bibinfo {author} {\bibfnamefont {V.~M.}\ \bibnamefont {Weaver}},\ }\bibfield  {title} {\bibinfo {title} {Balancing forces: architectural control of mechanotransduction},\ }\href@noop {} {\bibfield  {journal} {\bibinfo  {journal} {Nature Reviews Molecular Cell Biology}\ }\textbf {\bibinfo {volume} {12}},\ \bibinfo {pages} {308} (\bibinfo {year} {2011})}\BibitemShut {NoStop}%
\bibitem [{\citenamefont {Ridley}\ \emph {et~al.}(2003)\citenamefont {Ridley}, \citenamefont {Schwartz}, \citenamefont {Burridge}, \citenamefont {Firtel}, \citenamefont {Ginsberg}, \citenamefont {Borisy}, \citenamefont {Parsons},\ and\ \citenamefont {Horwitz}}]{ridley2003cell}%
  \BibitemOpen
  \bibfield  {author} {\bibinfo {author} {\bibfnamefont {A.~J.}\ \bibnamefont {Ridley}}, \bibinfo {author} {\bibfnamefont {M.~A.}\ \bibnamefont {Schwartz}}, \bibinfo {author} {\bibfnamefont {K.}~\bibnamefont {Burridge}}, \bibinfo {author} {\bibfnamefont {R.~A.}\ \bibnamefont {Firtel}}, \bibinfo {author} {\bibfnamefont {M.~H.}\ \bibnamefont {Ginsberg}}, \bibinfo {author} {\bibfnamefont {G.}~\bibnamefont {Borisy}}, \bibinfo {author} {\bibfnamefont {J.~T.}\ \bibnamefont {Parsons}},\ and\ \bibinfo {author} {\bibfnamefont {A.~R.}\ \bibnamefont {Horwitz}},\ }\bibfield  {title} {\bibinfo {title} {Cell migration: integrating signals from front to back},\ }\href@noop {} {\bibfield  {journal} {\bibinfo  {journal} {Science}\ }\textbf {\bibinfo {volume} {302}},\ \bibinfo {pages} {1704} (\bibinfo {year} {2003})}\BibitemShut {NoStop}%
\bibitem [{\citenamefont {Jaalouk}\ and\ \citenamefont {Lammerding}(2009)}]{jaalouk2009mechanotransduction}%
  \BibitemOpen
  \bibfield  {author} {\bibinfo {author} {\bibfnamefont {D.~E.}\ \bibnamefont {Jaalouk}}\ and\ \bibinfo {author} {\bibfnamefont {J.}~\bibnamefont {Lammerding}},\ }\bibfield  {title} {\bibinfo {title} {Mechanotransduction gone awry},\ }\href@noop {} {\bibfield  {journal} {\bibinfo  {journal} {Nature Reviews Molecular Cell Biology}\ }\textbf {\bibinfo {volume} {10}},\ \bibinfo {pages} {63} (\bibinfo {year} {2009})}\BibitemShut {NoStop}%
\bibitem [{\citenamefont {Helmlinger}\ \emph {et~al.}(1997)\citenamefont {Helmlinger}, \citenamefont {Netti}, \citenamefont {Lichtenbeld}, \citenamefont {Melder},\ and\ \citenamefont {Jain}}]{helmlinger1997solid}%
  \BibitemOpen
  \bibfield  {author} {\bibinfo {author} {\bibfnamefont {G.}~\bibnamefont {Helmlinger}}, \bibinfo {author} {\bibfnamefont {P.~A.}\ \bibnamefont {Netti}}, \bibinfo {author} {\bibfnamefont {H.~C.}\ \bibnamefont {Lichtenbeld}}, \bibinfo {author} {\bibfnamefont {R.~J.}\ \bibnamefont {Melder}},\ and\ \bibinfo {author} {\bibfnamefont {R.~K.}\ \bibnamefont {Jain}},\ }\bibfield  {title} {\bibinfo {title} {Solid stress inhibits the growth of multicellular tumor spheroids},\ }\href@noop {} {\bibfield  {journal} {\bibinfo  {journal} {Nature Biotechnology}\ }\textbf {\bibinfo {volume} {15}},\ \bibinfo {pages} {778} (\bibinfo {year} {1997})}\BibitemShut {NoStop}%
\bibitem [{\citenamefont {Engler}\ \emph {et~al.}(2006)\citenamefont {Engler}, \citenamefont {Sen}, \citenamefont {Sweeney},\ and\ \citenamefont {Discher}}]{engler2006matrix}%
  \BibitemOpen
  \bibfield  {author} {\bibinfo {author} {\bibfnamefont {A.~J.}\ \bibnamefont {Engler}}, \bibinfo {author} {\bibfnamefont {S.}~\bibnamefont {Sen}}, \bibinfo {author} {\bibfnamefont {H.~L.}\ \bibnamefont {Sweeney}},\ and\ \bibinfo {author} {\bibfnamefont {D.~E.}\ \bibnamefont {Discher}},\ }\bibfield  {title} {\bibinfo {title} {Matrix elasticity directs stem cell lineage specification},\ }\href@noop {} {\bibfield  {journal} {\bibinfo  {journal} {Cell}\ }\textbf {\bibinfo {volume} {126}},\ \bibinfo {pages} {677} (\bibinfo {year} {2006})}\BibitemShut {NoStop}%
\bibitem [{\citenamefont {Wang}\ \emph {et~al.}(2009)\citenamefont {Wang}, \citenamefont {Tytell},\ and\ \citenamefont {Ingber}}]{wang2009mechanotransduction}%
  \BibitemOpen
  \bibfield  {author} {\bibinfo {author} {\bibfnamefont {N.}~\bibnamefont {Wang}}, \bibinfo {author} {\bibfnamefont {J.~D.}\ \bibnamefont {Tytell}},\ and\ \bibinfo {author} {\bibfnamefont {D.~E.}\ \bibnamefont {Ingber}},\ }\bibfield  {title} {\bibinfo {title} {Mechanotransduction at a distance: mechanically coupling the extracellular matrix with the nucleus},\ }\href@noop {} {\bibfield  {journal} {\bibinfo  {journal} {Nature Reviews Molecular Cell Biology}\ }\textbf {\bibinfo {volume} {10}},\ \bibinfo {pages} {75} (\bibinfo {year} {2009})}\BibitemShut {NoStop}%
\bibitem [{\citenamefont {Baker}\ \emph {et~al.}(2009)\citenamefont {Baker}, \citenamefont {Bonnecaze},\ and\ \citenamefont {Zaman}}]{baker2009extracellular}%
  \BibitemOpen
  \bibfield  {author} {\bibinfo {author} {\bibfnamefont {E.~L.}\ \bibnamefont {Baker}}, \bibinfo {author} {\bibfnamefont {R.~T.}\ \bibnamefont {Bonnecaze}},\ and\ \bibinfo {author} {\bibfnamefont {M.~H.}\ \bibnamefont {Zaman}},\ }\bibfield  {title} {\bibinfo {title} {Extracellular matrix stiffness and architecture govern intracellular rheology in cancer},\ }\href@noop {} {\bibfield  {journal} {\bibinfo  {journal} {Biophysical Journal}\ }\textbf {\bibinfo {volume} {97}},\ \bibinfo {pages} {1013} (\bibinfo {year} {2009})}\BibitemShut {NoStop}%
\bibitem [{\citenamefont {Chaudhuri}\ \emph {et~al.}(2020)\citenamefont {Chaudhuri}, \citenamefont {Cooper-White}, \citenamefont {Janmey}, \citenamefont {Mooney},\ and\ \citenamefont {Shenoy}}]{chaudhuri2020effects}%
  \BibitemOpen
  \bibfield  {author} {\bibinfo {author} {\bibfnamefont {O.}~\bibnamefont {Chaudhuri}}, \bibinfo {author} {\bibfnamefont {J.}~\bibnamefont {Cooper-White}}, \bibinfo {author} {\bibfnamefont {P.~A.}\ \bibnamefont {Janmey}}, \bibinfo {author} {\bibfnamefont {D.~J.}\ \bibnamefont {Mooney}},\ and\ \bibinfo {author} {\bibfnamefont {V.~B.}\ \bibnamefont {Shenoy}},\ }\bibfield  {title} {\bibinfo {title} {Effects of extracellular matrix viscoelasticity on cellular behaviour},\ }\href@noop {} {\bibfield  {journal} {\bibinfo  {journal} {Nature}\ }\textbf {\bibinfo {volume} {584}},\ \bibinfo {pages} {535} (\bibinfo {year} {2020})}\BibitemShut {NoStop}%
\bibitem [{\citenamefont {Vining}\ and\ \citenamefont {Mooney}(2017)}]{vining2017mechanical}%
  \BibitemOpen
  \bibfield  {author} {\bibinfo {author} {\bibfnamefont {K.~H.}\ \bibnamefont {Vining}}\ and\ \bibinfo {author} {\bibfnamefont {D.~J.}\ \bibnamefont {Mooney}},\ }\bibfield  {title} {\bibinfo {title} {Mechanical forces direct stem cell behaviour in development and regeneration},\ }\href@noop {} {\bibfield  {journal} {\bibinfo  {journal} {Nature Reviews Molecular Cell Biology}\ }\textbf {\bibinfo {volume} {18}},\ \bibinfo {pages} {728} (\bibinfo {year} {2017})}\BibitemShut {NoStop}%
\bibitem [{\citenamefont {Jonietz}(2012)}]{jonietz2012mechanics}%
  \BibitemOpen
  \bibfield  {author} {\bibinfo {author} {\bibfnamefont {E.}~\bibnamefont {Jonietz}},\ }\bibfield  {title} {\bibinfo {title} {Mechanics: The forces of cancer},\ }\href@noop {} {\bibfield  {journal} {\bibinfo  {journal} {Nature}\ }\textbf {\bibinfo {volume} {491}},\ \bibinfo {pages} {S56} (\bibinfo {year} {2012})}\BibitemShut {NoStop}%
\bibitem [{\citenamefont {Troyanova-Wood}\ \emph {et~al.}(2019)\citenamefont {Troyanova-Wood}, \citenamefont {Meng},\ and\ \citenamefont {Yakovlev}}]{troyanova2019differentiating}%
  \BibitemOpen
  \bibfield  {author} {\bibinfo {author} {\bibfnamefont {M.}~\bibnamefont {Troyanova-Wood}}, \bibinfo {author} {\bibfnamefont {Z.}~\bibnamefont {Meng}},\ and\ \bibinfo {author} {\bibfnamefont {V.~V.}\ \bibnamefont {Yakovlev}},\ }\bibfield  {title} {\bibinfo {title} {Differentiating melanoma and healthy tissues based on elasticity-specific {B}rillouin microspectroscopy},\ }\href@noop {} {\bibfield  {journal} {\bibinfo  {journal} {Biomedical Optics Express}\ }\textbf {\bibinfo {volume} {10}},\ \bibinfo {pages} {1774} (\bibinfo {year} {2019})}\BibitemShut {NoStop}%
\bibitem [{\citenamefont {Lu}\ \emph {et~al.}(2012)\citenamefont {Lu}, \citenamefont {Weaver},\ and\ \citenamefont {Werb}}]{lu2012extracellular}%
  \BibitemOpen
  \bibfield  {author} {\bibinfo {author} {\bibfnamefont {P.}~\bibnamefont {Lu}}, \bibinfo {author} {\bibfnamefont {V.~M.}\ \bibnamefont {Weaver}},\ and\ \bibinfo {author} {\bibfnamefont {Z.}~\bibnamefont {Werb}},\ }\bibfield  {title} {\bibinfo {title} {The extracellular matrix: a dynamic niche in cancer progression},\ }\href@noop {} {\bibfield  {journal} {\bibinfo  {journal} {Journal of Cell Biology}\ }\textbf {\bibinfo {volume} {196}},\ \bibinfo {pages} {395} (\bibinfo {year} {2012})}\BibitemShut {NoStop}%
\bibitem [{\citenamefont {Friedl}(2004)}]{friedl2004dynamic}%
  \BibitemOpen
  \bibfield  {author} {\bibinfo {author} {\bibfnamefont {P.}~\bibnamefont {Friedl}},\ }\bibfield  {title} {\bibinfo {title} {Dynamic imaging of cellular interactions with extracellular matrix},\ }\href@noop {} {\bibfield  {journal} {\bibinfo  {journal} {Histochemistry and Cell Biology}\ }\textbf {\bibinfo {volume} {122}},\ \bibinfo {pages} {183} (\bibinfo {year} {2004})}\BibitemShut {NoStop}%
\bibitem [{\citenamefont {Friedl}(2019)}]{friedl2019rethinking}%
  \BibitemOpen
  \bibfield  {author} {\bibinfo {author} {\bibfnamefont {P.}~\bibnamefont {Friedl}},\ }\bibfield  {title} {\bibinfo {title} {Rethinking research into metastasis},\ }\href@noop {} {\bibfield  {journal} {\bibinfo  {journal} {Elife}\ }\textbf {\bibinfo {volume} {8}},\ \bibinfo {pages} {e53511} (\bibinfo {year} {2019})}\BibitemShut {NoStop}%
\bibitem [{\citenamefont {Friedl}\ and\ \citenamefont {Alexander}(2011)}]{friedl2011cancer}%
  \BibitemOpen
  \bibfield  {author} {\bibinfo {author} {\bibfnamefont {P.}~\bibnamefont {Friedl}}\ and\ \bibinfo {author} {\bibfnamefont {S.}~\bibnamefont {Alexander}},\ }\bibfield  {title} {\bibinfo {title} {Cancer invasion and the microenvironment: plasticity and reciprocity},\ }\href@noop {} {\bibfield  {journal} {\bibinfo  {journal} {Cell}\ }\textbf {\bibinfo {volume} {147}},\ \bibinfo {pages} {992} (\bibinfo {year} {2011})}\BibitemShut {NoStop}%
\bibitem [{\citenamefont {Bao}\ and\ \citenamefont {Suresh}(2003)}]{bao2003cell}%
  \BibitemOpen
  \bibfield  {author} {\bibinfo {author} {\bibfnamefont {G.}~\bibnamefont {Bao}}\ and\ \bibinfo {author} {\bibfnamefont {S.}~\bibnamefont {Suresh}},\ }\bibfield  {title} {\bibinfo {title} {Cell and molecular mechanics of biological materials},\ }\href@noop {} {\bibfield  {journal} {\bibinfo  {journal} {Nature Materials}\ }\textbf {\bibinfo {volume} {2}},\ \bibinfo {pages} {715} (\bibinfo {year} {2003})}\BibitemShut {NoStop}%
\bibitem [{\citenamefont {Savin}\ \emph {et~al.}(2011)\citenamefont {Savin}, \citenamefont {Kurpios}, \citenamefont {Shyer}, \citenamefont {Florescu}, \citenamefont {Liang}, \citenamefont {Mahadevan},\ and\ \citenamefont {Tabin}}]{savin2011growth}%
  \BibitemOpen
  \bibfield  {author} {\bibinfo {author} {\bibfnamefont {T.}~\bibnamefont {Savin}}, \bibinfo {author} {\bibfnamefont {N.~A.}\ \bibnamefont {Kurpios}}, \bibinfo {author} {\bibfnamefont {A.~E.}\ \bibnamefont {Shyer}}, \bibinfo {author} {\bibfnamefont {P.}~\bibnamefont {Florescu}}, \bibinfo {author} {\bibfnamefont {H.}~\bibnamefont {Liang}}, \bibinfo {author} {\bibfnamefont {L.}~\bibnamefont {Mahadevan}},\ and\ \bibinfo {author} {\bibfnamefont {C.~J.}\ \bibnamefont {Tabin}},\ }\bibfield  {title} {\bibinfo {title} {On the growth and form of the gut},\ }\href@noop {} {\bibfield  {journal} {\bibinfo  {journal} {Nature}\ }\textbf {\bibinfo {volume} {476}},\ \bibinfo {pages} {57} (\bibinfo {year} {2011})}\BibitemShut {NoStop}%
\bibitem [{\citenamefont {Blacher}\ and\ \citenamefont {Safar}(2005)}]{blacher2005large}%
  \BibitemOpen
  \bibfield  {author} {\bibinfo {author} {\bibfnamefont {J.}~\bibnamefont {Blacher}}\ and\ \bibinfo {author} {\bibfnamefont {M.~E.}\ \bibnamefont {Safar}},\ }\bibfield  {title} {\bibinfo {title} {Large-artery stiffness, hypertension and cardiovascular risk in older patients},\ }\href@noop {} {\bibfield  {journal} {\bibinfo  {journal} {Nature Clinical Practice Cardiovascular Medicine}\ }\textbf {\bibinfo {volume} {2}},\ \bibinfo {pages} {450} (\bibinfo {year} {2005})}\BibitemShut {NoStop}%
\bibitem [{\citenamefont {Oates}\ \emph {et~al.}(2009)\citenamefont {Oates}, \citenamefont {Gorfinkiel}, \citenamefont {Gonzalez-Gaitan},\ and\ \citenamefont {Heisenberg}}]{oates2009quantitative}%
  \BibitemOpen
  \bibfield  {author} {\bibinfo {author} {\bibfnamefont {A.~C.}\ \bibnamefont {Oates}}, \bibinfo {author} {\bibfnamefont {N.}~\bibnamefont {Gorfinkiel}}, \bibinfo {author} {\bibfnamefont {M.}~\bibnamefont {Gonzalez-Gaitan}},\ and\ \bibinfo {author} {\bibfnamefont {C.-P.}\ \bibnamefont {Heisenberg}},\ }\bibfield  {title} {\bibinfo {title} {Quantitative approaches in developmental biology},\ }\href@noop {} {\bibfield  {journal} {\bibinfo  {journal} {Nature Reviews Genetics}\ }\textbf {\bibinfo {volume} {10}},\ \bibinfo {pages} {517} (\bibinfo {year} {2009})}\BibitemShut {NoStop}%
\bibitem [{\citenamefont {Gillespie}\ and\ \citenamefont {Walker}(2001)}]{gillespie2001molecular}%
  \BibitemOpen
  \bibfield  {author} {\bibinfo {author} {\bibfnamefont {P.~G.}\ \bibnamefont {Gillespie}}\ and\ \bibinfo {author} {\bibfnamefont {R.~G.}\ \bibnamefont {Walker}},\ }\bibfield  {title} {\bibinfo {title} {Molecular basis of mechanosensory transduction},\ }\href@noop {} {\bibfield  {journal} {\bibinfo  {journal} {Nature}\ }\textbf {\bibinfo {volume} {413}},\ \bibinfo {pages} {194} (\bibinfo {year} {2001})}\BibitemShut {NoStop}%
\bibitem [{\citenamefont {Gaharwar}\ \emph {et~al.}(2020)\citenamefont {Gaharwar}, \citenamefont {Singh},\ and\ \citenamefont {Khademhosseini}}]{gaharwar2020engineered}%
  \BibitemOpen
  \bibfield  {author} {\bibinfo {author} {\bibfnamefont {A.~K.}\ \bibnamefont {Gaharwar}}, \bibinfo {author} {\bibfnamefont {I.}~\bibnamefont {Singh}},\ and\ \bibinfo {author} {\bibfnamefont {A.}~\bibnamefont {Khademhosseini}},\ }\bibfield  {title} {\bibinfo {title} {Engineered biomaterials for in situ tissue regeneration},\ }\href@noop {} {\bibfield  {journal} {\bibinfo  {journal} {Nature Reviews Materials}\ }\textbf {\bibinfo {volume} {5}},\ \bibinfo {pages} {686} (\bibinfo {year} {2020})}\BibitemShut {NoStop}%
\bibitem [{\citenamefont {Ghosh}\ and\ \citenamefont {Ingber}(2007)}]{ghosh2007micromechanical}%
  \BibitemOpen
  \bibfield  {author} {\bibinfo {author} {\bibfnamefont {K.}~\bibnamefont {Ghosh}}\ and\ \bibinfo {author} {\bibfnamefont {D.~E.}\ \bibnamefont {Ingber}},\ }\bibfield  {title} {\bibinfo {title} {Micromechanical control of cell and tissue development: implications for tissue engineering},\ }\href@noop {} {\bibfield  {journal} {\bibinfo  {journal} {Advanced Drug Delivery Reviews}\ }\textbf {\bibinfo {volume} {59}},\ \bibinfo {pages} {1306} (\bibinfo {year} {2007})}\BibitemShut {NoStop}%
\bibitem [{\citenamefont {Roos}\ \emph {et~al.}(2010)\citenamefont {Roos}, \citenamefont {Bruinsma},\ and\ \citenamefont {Wuite}}]{roos2010physical}%
  \BibitemOpen
  \bibfield  {author} {\bibinfo {author} {\bibfnamefont {W.}~\bibnamefont {Roos}}, \bibinfo {author} {\bibfnamefont {R.}~\bibnamefont {Bruinsma}},\ and\ \bibinfo {author} {\bibfnamefont {G.}~\bibnamefont {Wuite}},\ }\bibfield  {title} {\bibinfo {title} {Physical virology},\ }\href@noop {} {\bibfield  {journal} {\bibinfo  {journal} {Nature Physics}\ }\textbf {\bibinfo {volume} {6}},\ \bibinfo {pages} {733} (\bibinfo {year} {2010})}\BibitemShut {NoStop}%
\bibitem [{\citenamefont {Karampatzakis}\ \emph {et~al.}(2017)\citenamefont {Karampatzakis}, \citenamefont {Song}, \citenamefont {Allsopp}, \citenamefont {Filloux}, \citenamefont {Rice}, \citenamefont {Cohen}, \citenamefont {Wohland},\ and\ \citenamefont {T{\"o}r{\"o}k}}]{karampatzakis2017probing}%
  \BibitemOpen
  \bibfield  {author} {\bibinfo {author} {\bibfnamefont {A.}~\bibnamefont {Karampatzakis}}, \bibinfo {author} {\bibfnamefont {C.}~\bibnamefont {Song}}, \bibinfo {author} {\bibfnamefont {L.}~\bibnamefont {Allsopp}}, \bibinfo {author} {\bibfnamefont {A.}~\bibnamefont {Filloux}}, \bibinfo {author} {\bibfnamefont {S.}~\bibnamefont {Rice}}, \bibinfo {author} {\bibfnamefont {Y.}~\bibnamefont {Cohen}}, \bibinfo {author} {\bibfnamefont {T.}~\bibnamefont {Wohland}},\ and\ \bibinfo {author} {\bibfnamefont {P.}~\bibnamefont {T{\"o}r{\"o}k}},\ }\bibfield  {title} {\bibinfo {title} {Probing the internal micromechanical properties of pseudomonas aeruginosa biofilms by {B}rillouin imaging},\ }\href@noop {} {\bibfield  {journal} {\bibinfo  {journal} {npj Biofilms and Microbiomes}\ }\textbf {\bibinfo {volume} {3}},\ \bibinfo {pages} {20} (\bibinfo {year} {2017})}\BibitemShut {NoStop}%
\bibitem [{\citenamefont {Langer}\ and\ \citenamefont {Tirrell}(2004)}]{langer2004designing}%
  \BibitemOpen
  \bibfield  {author} {\bibinfo {author} {\bibfnamefont {R.}~\bibnamefont {Langer}}\ and\ \bibinfo {author} {\bibfnamefont {D.~A.}\ \bibnamefont {Tirrell}},\ }\bibfield  {title} {\bibinfo {title} {Designing materials for biology and medicine},\ }\href@noop {} {\bibfield  {journal} {\bibinfo  {journal} {Nature}\ }\textbf {\bibinfo {volume} {428}},\ \bibinfo {pages} {487} (\bibinfo {year} {2004})}\BibitemShut {NoStop}%
\bibitem [{\citenamefont {Koski}\ \emph {et~al.}(2013)\citenamefont {Koski}, \citenamefont {Akhenblit}, \citenamefont {McKiernan},\ and\ \citenamefont {Yarger}}]{koski2013non}%
  \BibitemOpen
  \bibfield  {author} {\bibinfo {author} {\bibfnamefont {K.~J.}\ \bibnamefont {Koski}}, \bibinfo {author} {\bibfnamefont {P.}~\bibnamefont {Akhenblit}}, \bibinfo {author} {\bibfnamefont {K.}~\bibnamefont {McKiernan}},\ and\ \bibinfo {author} {\bibfnamefont {J.~L.}\ \bibnamefont {Yarger}},\ }\bibfield  {title} {\bibinfo {title} {Non-invasive determination of the complete elastic moduli of spider silks},\ }\href@noop {} {\bibfield  {journal} {\bibinfo  {journal} {Nature Materials}\ }\textbf {\bibinfo {volume} {12}},\ \bibinfo {pages} {262} (\bibinfo {year} {2013})}\BibitemShut {NoStop}%
\bibitem [{\citenamefont {Balaban}\ \emph {et~al.}(2001)\citenamefont {Balaban}, \citenamefont {Schwarz}, \citenamefont {Riveline}, \citenamefont {Goichberg}, \citenamefont {Tzur}, \citenamefont {Sabanay}, \citenamefont {Mahalu}, \citenamefont {Safran}, \citenamefont {Bershadsky}, \citenamefont {Addadi} \emph {et~al.}}]{balaban2001force}%
  \BibitemOpen
  \bibfield  {author} {\bibinfo {author} {\bibfnamefont {N.~Q.}\ \bibnamefont {Balaban}}, \bibinfo {author} {\bibfnamefont {U.~S.}\ \bibnamefont {Schwarz}}, \bibinfo {author} {\bibfnamefont {D.}~\bibnamefont {Riveline}}, \bibinfo {author} {\bibfnamefont {P.}~\bibnamefont {Goichberg}}, \bibinfo {author} {\bibfnamefont {G.}~\bibnamefont {Tzur}}, \bibinfo {author} {\bibfnamefont {I.}~\bibnamefont {Sabanay}}, \bibinfo {author} {\bibfnamefont {D.}~\bibnamefont {Mahalu}}, \bibinfo {author} {\bibfnamefont {S.}~\bibnamefont {Safran}}, \bibinfo {author} {\bibfnamefont {A.}~\bibnamefont {Bershadsky}}, \bibinfo {author} {\bibfnamefont {L.}~\bibnamefont {Addadi}}, \emph {et~al.},\ }\bibfield  {title} {\bibinfo {title} {Force and focal adhesion assembly: a close relationship studied using elastic micropatterned substrates},\ }\href@noop {} {\bibfield  {journal} {\bibinfo  {journal} {Nature Cell Biology}\ }\textbf {\bibinfo {volume} {3}},\ \bibinfo {pages} {466} (\bibinfo {year} {2001})}\BibitemShut {NoStop}%
\bibitem [{\citenamefont {Greenleaf}\ \emph {et~al.}(2003)\citenamefont {Greenleaf}, \citenamefont {Fatemi},\ and\ \citenamefont {Insana}}]{greenleaf2003selected}%
  \BibitemOpen
  \bibfield  {author} {\bibinfo {author} {\bibfnamefont {J.~F.}\ \bibnamefont {Greenleaf}}, \bibinfo {author} {\bibfnamefont {M.}~\bibnamefont {Fatemi}},\ and\ \bibinfo {author} {\bibfnamefont {M.}~\bibnamefont {Insana}},\ }\bibfield  {title} {\bibinfo {title} {Selected methods for imaging elastic properties of biological tissues},\ }\href@noop {} {\bibfield  {journal} {\bibinfo  {journal} {Annual Review of Biomedical Engineering}\ }\textbf {\bibinfo {volume} {5}},\ \bibinfo {pages} {57} (\bibinfo {year} {2003})}\BibitemShut {NoStop}%
\bibitem [{\citenamefont {Discher}\ \emph {et~al.}(2009)\citenamefont {Discher}, \citenamefont {Dong}, \citenamefont {Fredberg}, \citenamefont {Guilak}, \citenamefont {Ingber}, \citenamefont {Janmey}, \citenamefont {Kamm}, \citenamefont {Schmid-Sch{\"o}nbein},\ and\ \citenamefont {Weinbaum}}]{discher2009biomechanics}%
  \BibitemOpen
  \bibfield  {author} {\bibinfo {author} {\bibfnamefont {D.}~\bibnamefont {Discher}}, \bibinfo {author} {\bibfnamefont {C.}~\bibnamefont {Dong}}, \bibinfo {author} {\bibfnamefont {J.~J.}\ \bibnamefont {Fredberg}}, \bibinfo {author} {\bibfnamefont {F.}~\bibnamefont {Guilak}}, \bibinfo {author} {\bibfnamefont {D.}~\bibnamefont {Ingber}}, \bibinfo {author} {\bibfnamefont {P.}~\bibnamefont {Janmey}}, \bibinfo {author} {\bibfnamefont {R.~D.}\ \bibnamefont {Kamm}}, \bibinfo {author} {\bibfnamefont {G.~W.}\ \bibnamefont {Schmid-Sch{\"o}nbein}},\ and\ \bibinfo {author} {\bibfnamefont {S.}~\bibnamefont {Weinbaum}},\ }\bibfield  {title} {\bibinfo {title} {Biomechanics: cell research and applications for the next decade},\ }\href@noop {} {\bibfield  {journal} {\bibinfo  {journal} {Annals of Biomedical Engineering}\ }\textbf {\bibinfo {volume} {37}},\ \bibinfo {pages} {847} (\bibinfo {year} {2009})}\BibitemShut {NoStop}%
\bibitem [{\citenamefont {Ingber}(2006)}]{ingber2006mechanical}%
  \BibitemOpen
  \bibfield  {author} {\bibinfo {author} {\bibfnamefont {D.~E.}\ \bibnamefont {Ingber}},\ }\bibfield  {title} {\bibinfo {title} {Mechanical control of tissue morphogenesis during embryological development},\ }\href@noop {} {\bibfield  {journal} {\bibinfo  {journal} {The International Journal of Developmental Biology}\ }\textbf {\bibinfo {volume} {50}},\ \bibinfo {pages} {255} (\bibinfo {year} {2006})}\BibitemShut {NoStop}%
\bibitem [{\citenamefont {Mouw}\ \emph {et~al.}(2014)\citenamefont {Mouw}, \citenamefont {Ou},\ and\ \citenamefont {Weaver}}]{mouw2014extracellular}%
  \BibitemOpen
  \bibfield  {author} {\bibinfo {author} {\bibfnamefont {J.~K.}\ \bibnamefont {Mouw}}, \bibinfo {author} {\bibfnamefont {G.}~\bibnamefont {Ou}},\ and\ \bibinfo {author} {\bibfnamefont {V.~M.}\ \bibnamefont {Weaver}},\ }\bibfield  {title} {\bibinfo {title} {Extracellular matrix assembly: a multiscale deconstruction},\ }\href@noop {} {\bibfield  {journal} {\bibinfo  {journal} {Nature Reviews Molecular Cell Biology}\ }\textbf {\bibinfo {volume} {15}},\ \bibinfo {pages} {771} (\bibinfo {year} {2014})}\BibitemShut {NoStop}%
\bibitem [{\citenamefont {List}\ \emph {et~al.}(2005)\citenamefont {List}, \citenamefont {Szabo}, \citenamefont {Molinolo}, \citenamefont {Sriuranpong}, \citenamefont {Redeye}, \citenamefont {Murdock}, \citenamefont {Burke}, \citenamefont {Nielsen}, \citenamefont {Gutkind},\ and\ \citenamefont {Bugge}}]{list2005deregulated}%
  \BibitemOpen
  \bibfield  {author} {\bibinfo {author} {\bibfnamefont {K.}~\bibnamefont {List}}, \bibinfo {author} {\bibfnamefont {R.}~\bibnamefont {Szabo}}, \bibinfo {author} {\bibfnamefont {A.}~\bibnamefont {Molinolo}}, \bibinfo {author} {\bibfnamefont {V.}~\bibnamefont {Sriuranpong}}, \bibinfo {author} {\bibfnamefont {V.}~\bibnamefont {Redeye}}, \bibinfo {author} {\bibfnamefont {T.}~\bibnamefont {Murdock}}, \bibinfo {author} {\bibfnamefont {B.}~\bibnamefont {Burke}}, \bibinfo {author} {\bibfnamefont {B.~S.}\ \bibnamefont {Nielsen}}, \bibinfo {author} {\bibfnamefont {J.~S.}\ \bibnamefont {Gutkind}},\ and\ \bibinfo {author} {\bibfnamefont {T.~H.}\ \bibnamefont {Bugge}},\ }\bibfield  {title} {\bibinfo {title} {Deregulated matriptase causes ras-independent multistage carcinogenesis and promotes ras-mediated malignant transformation},\ }\href@noop {} {\bibfield  {journal} {\bibinfo  {journal} {Genes \& Development}\ }\textbf {\bibinfo {volume} {19}},\ \bibinfo {pages} {1934} (\bibinfo {year} {2005})}\BibitemShut {NoStop}%
\bibitem [{\citenamefont {Paszek}\ \emph {et~al.}(2005)\citenamefont {Paszek}, \citenamefont {Zahir}, \citenamefont {Johnson}, \citenamefont {Lakins}, \citenamefont {Rozenberg}, \citenamefont {Gefen}, \citenamefont {Reinhart-King}, \citenamefont {Margulies}, \citenamefont {Dembo}, \citenamefont {Boettiger} \emph {et~al.}}]{paszek2005tensional}%
  \BibitemOpen
  \bibfield  {author} {\bibinfo {author} {\bibfnamefont {M.~J.}\ \bibnamefont {Paszek}}, \bibinfo {author} {\bibfnamefont {N.}~\bibnamefont {Zahir}}, \bibinfo {author} {\bibfnamefont {K.~R.}\ \bibnamefont {Johnson}}, \bibinfo {author} {\bibfnamefont {J.~N.}\ \bibnamefont {Lakins}}, \bibinfo {author} {\bibfnamefont {G.~I.}\ \bibnamefont {Rozenberg}}, \bibinfo {author} {\bibfnamefont {A.}~\bibnamefont {Gefen}}, \bibinfo {author} {\bibfnamefont {C.~A.}\ \bibnamefont {Reinhart-King}}, \bibinfo {author} {\bibfnamefont {S.~S.}\ \bibnamefont {Margulies}}, \bibinfo {author} {\bibfnamefont {M.}~\bibnamefont {Dembo}}, \bibinfo {author} {\bibfnamefont {D.}~\bibnamefont {Boettiger}}, \emph {et~al.},\ }\bibfield  {title} {\bibinfo {title} {Tensional homeostasis and the malignant phenotype},\ }\href@noop {} {\bibfield  {journal} {\bibinfo  {journal} {Cancer Cell}\ }\textbf {\bibinfo {volume} {8}},\ \bibinfo {pages} {241} (\bibinfo {year} {2005})}\BibitemShut {NoStop}%
\bibitem [{\citenamefont {Radisky}\ \emph {et~al.}(2005)\citenamefont {Radisky}, \citenamefont {Levy}, \citenamefont {Littlepage}, \citenamefont {Liu}, \citenamefont {Nelson}, \citenamefont {Fata}, \citenamefont {Leake}, \citenamefont {Godden}, \citenamefont {Albertson}, \citenamefont {Angela~Nieto} \emph {et~al.}}]{radisky2005rac1b}%
  \BibitemOpen
  \bibfield  {author} {\bibinfo {author} {\bibfnamefont {D.~C.}\ \bibnamefont {Radisky}}, \bibinfo {author} {\bibfnamefont {D.~D.}\ \bibnamefont {Levy}}, \bibinfo {author} {\bibfnamefont {L.~E.}\ \bibnamefont {Littlepage}}, \bibinfo {author} {\bibfnamefont {H.}~\bibnamefont {Liu}}, \bibinfo {author} {\bibfnamefont {C.~M.}\ \bibnamefont {Nelson}}, \bibinfo {author} {\bibfnamefont {J.~E.}\ \bibnamefont {Fata}}, \bibinfo {author} {\bibfnamefont {D.}~\bibnamefont {Leake}}, \bibinfo {author} {\bibfnamefont {E.~L.}\ \bibnamefont {Godden}}, \bibinfo {author} {\bibfnamefont {D.~G.}\ \bibnamefont {Albertson}}, \bibinfo {author} {\bibfnamefont {M.}~\bibnamefont {Angela~Nieto}}, \emph {et~al.},\ }\bibfield  {title} {\bibinfo {title} {Rac1b and reactive oxygen species mediate mmp-3-induced emt and genomic instability},\ }\href@noop {} {\bibfield  {journal} {\bibinfo  {journal} {Nature}\ }\textbf {\bibinfo {volume} {436}},\ \bibinfo {pages} {123} (\bibinfo {year} {2005})}\BibitemShut {NoStop}%
\bibitem [{\citenamefont {Suresh}(2007)}]{suresh2007biomechanics}%
  \BibitemOpen
  \bibfield  {author} {\bibinfo {author} {\bibfnamefont {S.}~\bibnamefont {Suresh}},\ }\bibfield  {title} {\bibinfo {title} {Biomechanics and biophysics of cancer cells},\ }\href@noop {} {\bibfield  {journal} {\bibinfo  {journal} {Acta Biomaterialia}\ }\textbf {\bibinfo {volume} {3}},\ \bibinfo {pages} {413} (\bibinfo {year} {2007})}\BibitemShut {NoStop}%
\bibitem [{\citenamefont {Huwart}\ \emph {et~al.}(2008)\citenamefont {Huwart}, \citenamefont {Sempoux}, \citenamefont {Vicaut}, \citenamefont {Salameh}, \citenamefont {Annet}, \citenamefont {Danse}, \citenamefont {Peeters}, \citenamefont {ter Beek}, \citenamefont {Rahier}, \citenamefont {Sinkus} \emph {et~al.}}]{huwart2008magnetic}%
  \BibitemOpen
  \bibfield  {author} {\bibinfo {author} {\bibfnamefont {L.}~\bibnamefont {Huwart}}, \bibinfo {author} {\bibfnamefont {C.}~\bibnamefont {Sempoux}}, \bibinfo {author} {\bibfnamefont {E.}~\bibnamefont {Vicaut}}, \bibinfo {author} {\bibfnamefont {N.}~\bibnamefont {Salameh}}, \bibinfo {author} {\bibfnamefont {L.}~\bibnamefont {Annet}}, \bibinfo {author} {\bibfnamefont {E.}~\bibnamefont {Danse}}, \bibinfo {author} {\bibfnamefont {F.}~\bibnamefont {Peeters}}, \bibinfo {author} {\bibfnamefont {L.~C.}\ \bibnamefont {ter Beek}}, \bibinfo {author} {\bibfnamefont {J.}~\bibnamefont {Rahier}}, \bibinfo {author} {\bibfnamefont {R.}~\bibnamefont {Sinkus}}, \emph {et~al.},\ }\bibfield  {title} {\bibinfo {title} {Magnetic resonance elastography for the noninvasive staging of liver fibrosis},\ }\href@noop {} {\bibfield  {journal} {\bibinfo  {journal} {Gastroenterology}\ }\textbf {\bibinfo {volume} {135}},\ \bibinfo {pages} {32} (\bibinfo {year} {2008})}\BibitemShut {NoStop}%
\bibitem [{\citenamefont {Butcher}\ \emph {et~al.}(2009)\citenamefont {Butcher}, \citenamefont {Alliston},\ and\ \citenamefont {Weaver}}]{butcher2009tense}%
  \BibitemOpen
  \bibfield  {author} {\bibinfo {author} {\bibfnamefont {D.~T.}\ \bibnamefont {Butcher}}, \bibinfo {author} {\bibfnamefont {T.}~\bibnamefont {Alliston}},\ and\ \bibinfo {author} {\bibfnamefont {V.~M.}\ \bibnamefont {Weaver}},\ }\bibfield  {title} {\bibinfo {title} {A tense situation: forcing tumour progression},\ }\href@noop {} {\bibfield  {journal} {\bibinfo  {journal} {Nature Reviews Cancer}\ }\textbf {\bibinfo {volume} {9}},\ \bibinfo {pages} {108} (\bibinfo {year} {2009})}\BibitemShut {NoStop}%
\bibitem [{\citenamefont {Scarcelli}\ \emph {et~al.}(2012)\citenamefont {Scarcelli}, \citenamefont {Pineda},\ and\ \citenamefont {Yun}}]{scarcelli2012brillouin}%
  \BibitemOpen
  \bibfield  {author} {\bibinfo {author} {\bibfnamefont {G.}~\bibnamefont {Scarcelli}}, \bibinfo {author} {\bibfnamefont {R.}~\bibnamefont {Pineda}},\ and\ \bibinfo {author} {\bibfnamefont {S.~H.}\ \bibnamefont {Yun}},\ }\bibfield  {title} {\bibinfo {title} {Brillouin optical microscopy for corneal biomechanics},\ }\href@noop {} {\bibfield  {journal} {\bibinfo  {journal} {Investigative Ophthalmology \& Visual Science}\ }\textbf {\bibinfo {volume} {53}},\ \bibinfo {pages} {185} (\bibinfo {year} {2012})}\BibitemShut {NoStop}%
\bibitem [{\citenamefont {Stachs}\ \emph {et~al.}(2012)\citenamefont {Stachs}, \citenamefont {Rei{\ss}}, \citenamefont {Guthoff},\ and\ \citenamefont {Stolz}}]{stachs2012spatially}%
  \BibitemOpen
  \bibfield  {author} {\bibinfo {author} {\bibfnamefont {O.}~\bibnamefont {Stachs}}, \bibinfo {author} {\bibfnamefont {S.}~\bibnamefont {Rei{\ss}}}, \bibinfo {author} {\bibfnamefont {R.}~\bibnamefont {Guthoff}},\ and\ \bibinfo {author} {\bibfnamefont {H.}~\bibnamefont {Stolz}},\ }\bibfield  {title} {\bibinfo {title} {Spatially resolved {B}rillouin spectroscopy for in vivo determination of the biomechanical properties of the crystalline lenses},\ }in\ \href@noop {} {\emph {\bibinfo {booktitle} {Ophthalmic Technologies XXII}}},\ Vol.\ \bibinfo {volume} {8209}\ (\bibinfo {organization} {SPIE},\ \bibinfo {year} {2012})\ pp.\ \bibinfo {pages} {73--78}\BibitemShut {NoStop}%
\bibitem [{\citenamefont {Scarcelli}\ \emph {et~al.}(2013)\citenamefont {Scarcelli}, \citenamefont {Kling}, \citenamefont {Quijano}, \citenamefont {Pineda}, \citenamefont {Marcos},\ and\ \citenamefont {Yun}}]{scarcelli2013brillouin}%
  \BibitemOpen
  \bibfield  {author} {\bibinfo {author} {\bibfnamefont {G.}~\bibnamefont {Scarcelli}}, \bibinfo {author} {\bibfnamefont {S.}~\bibnamefont {Kling}}, \bibinfo {author} {\bibfnamefont {E.}~\bibnamefont {Quijano}}, \bibinfo {author} {\bibfnamefont {R.}~\bibnamefont {Pineda}}, \bibinfo {author} {\bibfnamefont {S.}~\bibnamefont {Marcos}},\ and\ \bibinfo {author} {\bibfnamefont {S.~H.}\ \bibnamefont {Yun}},\ }\bibfield  {title} {\bibinfo {title} {Brillouin microscopy of collagen crosslinking: noncontact depth-dependent analysis of corneal elastic modulus},\ }\href@noop {} {\bibfield  {journal} {\bibinfo  {journal} {Investigative Ophthalmology \& Visual Science}\ }\textbf {\bibinfo {volume} {54}},\ \bibinfo {pages} {1418} (\bibinfo {year} {2013})}\BibitemShut {NoStop}%
\bibitem [{\citenamefont {Comoglio}\ and\ \citenamefont {Trusolino}(2005)}]{comoglio2005cancer}%
  \BibitemOpen
  \bibfield  {author} {\bibinfo {author} {\bibfnamefont {P.~M.}\ \bibnamefont {Comoglio}}\ and\ \bibinfo {author} {\bibfnamefont {L.}~\bibnamefont {Trusolino}},\ }\bibfield  {title} {\bibinfo {title} {Cancer: the matrix is now in control},\ }\href@noop {} {\bibfield  {journal} {\bibinfo  {journal} {Nature Medicine}\ }\textbf {\bibinfo {volume} {11}},\ \bibinfo {pages} {1156} (\bibinfo {year} {2005})}\BibitemShut {NoStop}%
\bibitem [{\citenamefont {Spedden}\ and\ \citenamefont {Staii}(2013)}]{spedden2013neuron}%
  \BibitemOpen
  \bibfield  {author} {\bibinfo {author} {\bibfnamefont {E.}~\bibnamefont {Spedden}}\ and\ \bibinfo {author} {\bibfnamefont {C.}~\bibnamefont {Staii}},\ }\bibfield  {title} {\bibinfo {title} {Neuron biomechanics probed by atomic force microscopy},\ }\href@noop {} {\bibfield  {journal} {\bibinfo  {journal} {International Journal of Molecular Sciences}\ }\textbf {\bibinfo {volume} {14}},\ \bibinfo {pages} {16124} (\bibinfo {year} {2013})}\BibitemShut {NoStop}%
\bibitem [{\citenamefont {Van~Vliet}\ \emph {et~al.}(2003)\citenamefont {Van~Vliet}, \citenamefont {Bao},\ and\ \citenamefont {Suresh}}]{van2003biomechanics}%
  \BibitemOpen
  \bibfield  {author} {\bibinfo {author} {\bibfnamefont {K.}~\bibnamefont {Van~Vliet}}, \bibinfo {author} {\bibfnamefont {G.}~\bibnamefont {Bao}},\ and\ \bibinfo {author} {\bibfnamefont {S.}~\bibnamefont {Suresh}},\ }\bibfield  {title} {\bibinfo {title} {The biomechanics toolbox: experimental approaches for living cells and biomolecules},\ }\href@noop {} {\bibfield  {journal} {\bibinfo  {journal} {Acta Materialia}\ }\textbf {\bibinfo {volume} {51}},\ \bibinfo {pages} {5881} (\bibinfo {year} {2003})}\BibitemShut {NoStop}%
\bibitem [{\citenamefont {Randall}\ and\ \citenamefont {Vaughan}(1979)}]{randall1979brillouin}%
  \BibitemOpen
  \bibfield  {author} {\bibinfo {author} {\bibfnamefont {J.~T.}\ \bibnamefont {Randall}}\ and\ \bibinfo {author} {\bibfnamefont {J.~M.}\ \bibnamefont {Vaughan}},\ }\bibfield  {title} {\bibinfo {title} {Brillouin scattering in systems of biological significance},\ }\href@noop {} {\bibfield  {journal} {\bibinfo  {journal} {Philosophical Transactions of the Royal Society of London. Series A, Mathematical and Physical Sciences}\ }\textbf {\bibinfo {volume} {293}},\ \bibinfo {pages} {341} (\bibinfo {year} {1979})}\BibitemShut {NoStop}%
\bibitem [{\citenamefont {Editors}(2022)}]{Guardian}%
  \BibitemOpen
  \bibfield  {author} {\bibinfo {author} {\bibnamefont {Editors}},\ }\href@noop {} {\emph {\bibinfo {title} {The 10 biggest science stories of 2022 – chosen by scientists}}}\ (\bibinfo  {publisher} {The Guardian},\ \bibinfo {year} {2022})\BibitemShut {NoStop}%
\bibitem [{\citenamefont {Prevedel}\ \emph {et~al.}(2019)\citenamefont {Prevedel}, \citenamefont {Diz-Mu{\~n}oz}, \citenamefont {Ruocco},\ and\ \citenamefont {Antonacci}}]{prevedel2019brillouin}%
  \BibitemOpen
  \bibfield  {author} {\bibinfo {author} {\bibfnamefont {R.}~\bibnamefont {Prevedel}}, \bibinfo {author} {\bibfnamefont {A.}~\bibnamefont {Diz-Mu{\~n}oz}}, \bibinfo {author} {\bibfnamefont {G.}~\bibnamefont {Ruocco}},\ and\ \bibinfo {author} {\bibfnamefont {G.}~\bibnamefont {Antonacci}},\ }\bibfield  {title} {\bibinfo {title} {Brillouin microscopy: an emerging tool for mechanobiology},\ }\href@noop {} {\bibfield  {journal} {\bibinfo  {journal} {Nature Methods}\ }\textbf {\bibinfo {volume} {16}},\ \bibinfo {pages} {969} (\bibinfo {year} {2019})}\BibitemShut {NoStop}%
\bibitem [{\citenamefont {Poon}\ \emph {et~al.}(2020)\citenamefont {Poon}, \citenamefont {Chou}, \citenamefont {Cortie},\ and\ \citenamefont {Kabakova}}]{poon2020brillouin}%
  \BibitemOpen
  \bibfield  {author} {\bibinfo {author} {\bibfnamefont {C.}~\bibnamefont {Poon}}, \bibinfo {author} {\bibfnamefont {J.}~\bibnamefont {Chou}}, \bibinfo {author} {\bibfnamefont {M.}~\bibnamefont {Cortie}},\ and\ \bibinfo {author} {\bibfnamefont {I.}~\bibnamefont {Kabakova}},\ }\bibfield  {title} {\bibinfo {title} {Brillouin imaging for studies of micromechanics in biology and biomedicine: From current state-of-the-art to future clinical translation},\ }\href@noop {} {\bibfield  {journal} {\bibinfo  {journal} {Journal of Physics: Photonics}\ }\textbf {\bibinfo {volume} {3}},\ \bibinfo {pages} {012002} (\bibinfo {year} {2020})}\BibitemShut {NoStop}%
\bibitem [{\citenamefont {Palombo}\ and\ \citenamefont {Fioretto}(2019)}]{palombo2019brillouin}%
  \BibitemOpen
  \bibfield  {author} {\bibinfo {author} {\bibfnamefont {F.}~\bibnamefont {Palombo}}\ and\ \bibinfo {author} {\bibfnamefont {D.}~\bibnamefont {Fioretto}},\ }\bibfield  {title} {\bibinfo {title} {Brillouin light scattering: applications in biomedical sciences},\ }\href@noop {} {\bibfield  {journal} {\bibinfo  {journal} {Chemical Reviews}\ }\textbf {\bibinfo {volume} {119}},\ \bibinfo {pages} {7833} (\bibinfo {year} {2019})}\BibitemShut {NoStop}%
\bibitem [{\citenamefont {Elsayad}\ \emph {et~al.}(2019)\citenamefont {Elsayad}, \citenamefont {Palombo}, \citenamefont {Dehoux},\ and\ \citenamefont {Fioretto}}]{elsayad2019brillouin}%
  \BibitemOpen
  \bibfield  {author} {\bibinfo {author} {\bibfnamefont {K.}~\bibnamefont {Elsayad}}, \bibinfo {author} {\bibfnamefont {F.}~\bibnamefont {Palombo}}, \bibinfo {author} {\bibfnamefont {T.}~\bibnamefont {Dehoux}},\ and\ \bibinfo {author} {\bibfnamefont {D.}~\bibnamefont {Fioretto}},\ }\bibfield  {title} {\bibinfo {title} {Brillouin light scattering microspectroscopy for biomedical research and applications: introduction to feature issue},\ }\href@noop {} {\bibfield  {journal} {\bibinfo  {journal} {Biomedical Optics Express}\ }\textbf {\bibinfo {volume} {10}},\ \bibinfo {pages} {2670} (\bibinfo {year} {2019})}\BibitemShut {NoStop}%
\bibitem [{\citenamefont {Singaraju}\ \emph {et~al.}(2019)\citenamefont {Singaraju}, \citenamefont {Bahl},\ and\ \citenamefont {Stevens}}]{singaraju2019brillouin}%
  \BibitemOpen
  \bibfield  {author} {\bibinfo {author} {\bibfnamefont {A.~B.}\ \bibnamefont {Singaraju}}, \bibinfo {author} {\bibfnamefont {D.}~\bibnamefont {Bahl}},\ and\ \bibinfo {author} {\bibfnamefont {L.~L.}\ \bibnamefont {Stevens}},\ }\bibfield  {title} {\bibinfo {title} {Brillouin light scattering: development of a near century-old technique for characterizing the mechanical properties of materials},\ }\href@noop {} {\bibfield  {journal} {\bibinfo  {journal} {AAPS PharmSciTech}\ }\textbf {\bibinfo {volume} {20}},\ \bibinfo {pages} {109} (\bibinfo {year} {2019})}\BibitemShut {NoStop}%
\bibitem [{\citenamefont {Sandercock}(2005)}]{sandercock2005trends}%
  \BibitemOpen
  \bibfield  {author} {\bibinfo {author} {\bibfnamefont {J.}~\bibnamefont {Sandercock}},\ }\bibfield  {title} {\bibinfo {title} {Trends in {B}rillouin scattering: studies of opaque materials, supported films, and central modes},\ }\href@noop {} {\bibfield  {journal} {\bibinfo  {journal} {Light Scattering in Solids III: Recent Results}\ ,\ \bibinfo {pages} {173}} (\bibinfo {year} {2005})}\BibitemShut {NoStop}%
\bibitem [{\citenamefont {Dil}(1982)}]{dil1982brillouin}%
  \BibitemOpen
  \bibfield  {author} {\bibinfo {author} {\bibfnamefont {J.}~\bibnamefont {Dil}},\ }\bibfield  {title} {\bibinfo {title} {Brillouin scattering in condensed matter},\ }\href@noop {} {\bibfield  {journal} {\bibinfo  {journal} {Reports on Progress in Physics}\ }\textbf {\bibinfo {volume} {45}},\ \bibinfo {pages} {285} (\bibinfo {year} {1982})}\BibitemShut {NoStop}%
\bibitem [{\citenamefont {Scarcelli}\ \emph {et~al.}(2015)\citenamefont {Scarcelli}, \citenamefont {Polacheck}, \citenamefont {Nia}, \citenamefont {Patel}, \citenamefont {Grodzinsky}, \citenamefont {Kamm},\ and\ \citenamefont {Yun}}]{scarcelli2015noncontact}%
  \BibitemOpen
  \bibfield  {author} {\bibinfo {author} {\bibfnamefont {G.}~\bibnamefont {Scarcelli}}, \bibinfo {author} {\bibfnamefont {W.~J.}\ \bibnamefont {Polacheck}}, \bibinfo {author} {\bibfnamefont {H.~T.}\ \bibnamefont {Nia}}, \bibinfo {author} {\bibfnamefont {K.}~\bibnamefont {Patel}}, \bibinfo {author} {\bibfnamefont {A.~J.}\ \bibnamefont {Grodzinsky}}, \bibinfo {author} {\bibfnamefont {R.~D.}\ \bibnamefont {Kamm}},\ and\ \bibinfo {author} {\bibfnamefont {S.~H.}\ \bibnamefont {Yun}},\ }\bibfield  {title} {\bibinfo {title} {Noncontact three-dimensional mapping of intracellular hydromechanical properties by {B}rillouin microscopy},\ }\href@noop {} {\bibfield  {journal} {\bibinfo  {journal} {Nature Methods}\ }\textbf {\bibinfo {volume} {12}},\ \bibinfo {pages} {1132} (\bibinfo {year} {2015})}\BibitemShut {NoStop}%
\bibitem [{\citenamefont {Meng}\ \emph {et~al.}(2016)\citenamefont {Meng}, \citenamefont {Traverso}, \citenamefont {Ballmann}, \citenamefont {Troyanova-Wood},\ and\ \citenamefont {Yakovlev}}]{meng2016seeing}%
  \BibitemOpen
  \bibfield  {author} {\bibinfo {author} {\bibfnamefont {Z.}~\bibnamefont {Meng}}, \bibinfo {author} {\bibfnamefont {A.~J.}\ \bibnamefont {Traverso}}, \bibinfo {author} {\bibfnamefont {C.~W.}\ \bibnamefont {Ballmann}}, \bibinfo {author} {\bibfnamefont {M.~A.}\ \bibnamefont {Troyanova-Wood}},\ and\ \bibinfo {author} {\bibfnamefont {V.~V.}\ \bibnamefont {Yakovlev}},\ }\bibfield  {title} {\bibinfo {title} {Seeing cells in a new light: a renaissance of {B}rillouin spectroscopy},\ }\href@noop {} {\bibfield  {journal} {\bibinfo  {journal} {Advances in Optics and Photonics}\ }\textbf {\bibinfo {volume} {8}},\ \bibinfo {pages} {300} (\bibinfo {year} {2016})}\BibitemShut {NoStop}%
\bibitem [{\citenamefont {Kennedy}\ \emph {et~al.}(2017)\citenamefont {Kennedy}, \citenamefont {Wijesinghe},\ and\ \citenamefont {Sampson}}]{kennedy2017emergence}%
  \BibitemOpen
  \bibfield  {author} {\bibinfo {author} {\bibfnamefont {B.~F.}\ \bibnamefont {Kennedy}}, \bibinfo {author} {\bibfnamefont {P.}~\bibnamefont {Wijesinghe}},\ and\ \bibinfo {author} {\bibfnamefont {D.~D.}\ \bibnamefont {Sampson}},\ }\bibfield  {title} {\bibinfo {title} {The emergence of optical elastography in biomedicine},\ }\href@noop {} {\bibfield  {journal} {\bibinfo  {journal} {Nature Photonics}\ }\textbf {\bibinfo {volume} {11}},\ \bibinfo {pages} {215} (\bibinfo {year} {2017})}\BibitemShut {NoStop}%
\bibitem [{\citenamefont {Hawkins}\ and\ \citenamefont {Abrahamse}(2006)}]{hawkins2006role}%
  \BibitemOpen
  \bibfield  {author} {\bibinfo {author} {\bibfnamefont {D.~H.}\ \bibnamefont {Hawkins}}\ and\ \bibinfo {author} {\bibfnamefont {H.}~\bibnamefont {Abrahamse}},\ }\bibfield  {title} {\bibinfo {title} {The role of laser fluence in cell viability, proliferation, and membrane integrity of wounded human skin fibroblasts following helium-neon laser irradiation},\ }\href@noop {} {\bibfield  {journal} {\bibinfo  {journal} {Lasers in Surgery and Medicine: The Official Journal of the American Society for Laser Medicine and Surgery}\ }\textbf {\bibinfo {volume} {38}},\ \bibinfo {pages} {74} (\bibinfo {year} {2006})}\BibitemShut {NoStop}%
\bibitem [{\citenamefont {Denton}\ \emph {et~al.}(2006)\citenamefont {Denton}, \citenamefont {Foltz}, \citenamefont {Estlack}, \citenamefont {Stolarski}, \citenamefont {Noojin}, \citenamefont {Thomas}, \citenamefont {Eikum},\ and\ \citenamefont {Rockwell}}]{denton2006damage}%
  \BibitemOpen
  \bibfield  {author} {\bibinfo {author} {\bibfnamefont {M.~L.}\ \bibnamefont {Denton}}, \bibinfo {author} {\bibfnamefont {M.~S.}\ \bibnamefont {Foltz}}, \bibinfo {author} {\bibfnamefont {L.~E.}\ \bibnamefont {Estlack}}, \bibinfo {author} {\bibfnamefont {D.~J.}\ \bibnamefont {Stolarski}}, \bibinfo {author} {\bibfnamefont {G.~D.}\ \bibnamefont {Noojin}}, \bibinfo {author} {\bibfnamefont {R.~J.}\ \bibnamefont {Thomas}}, \bibinfo {author} {\bibfnamefont {D.}~\bibnamefont {Eikum}},\ and\ \bibinfo {author} {\bibfnamefont {B.~A.}\ \bibnamefont {Rockwell}},\ }\bibfield  {title} {\bibinfo {title} {Damage thresholds for exposure to nir and blue lasers in an in vitro rpe cell system},\ }\href@noop {} {\bibfield  {journal} {\bibinfo  {journal} {Investigative Ophthalmology \& Visual Science}\ }\textbf {\bibinfo {volume} {47}},\ \bibinfo {pages} {3065} (\bibinfo {year} {2006})}\BibitemShut {NoStop}%
\bibitem [{\citenamefont {Bixler}\ \emph {et~al.}(2014)\citenamefont {Bixler}, \citenamefont {Hokr}, \citenamefont {Denton}, \citenamefont {Noojin}, \citenamefont {Shingledecker}, \citenamefont {Beier}, \citenamefont {Thomas}, \citenamefont {Rockwell},\ and\ \citenamefont {Yakovlev}}]{bixler2014assessment}%
  \BibitemOpen
  \bibfield  {author} {\bibinfo {author} {\bibfnamefont {J.~N.}\ \bibnamefont {Bixler}}, \bibinfo {author} {\bibfnamefont {B.~H.}\ \bibnamefont {Hokr}}, \bibinfo {author} {\bibfnamefont {M.~L.}\ \bibnamefont {Denton}}, \bibinfo {author} {\bibfnamefont {G.~D.}\ \bibnamefont {Noojin}}, \bibinfo {author} {\bibfnamefont {A.~D.}\ \bibnamefont {Shingledecker}}, \bibinfo {author} {\bibfnamefont {H.~T.}\ \bibnamefont {Beier}}, \bibinfo {author} {\bibfnamefont {R.~J.}\ \bibnamefont {Thomas}}, \bibinfo {author} {\bibfnamefont {B.~A.}\ \bibnamefont {Rockwell}},\ and\ \bibinfo {author} {\bibfnamefont {V.~V.}\ \bibnamefont {Yakovlev}},\ }\bibfield  {title} {\bibinfo {title} {Assessment of tissue heating under tunable near-infrared radiation},\ }\href@noop {} {\bibfield  {journal} {\bibinfo  {journal} {Journal of Biomedical Optics}\ }\textbf {\bibinfo {volume} {19}},\ \bibinfo {pages} {070501} (\bibinfo {year} {2014})}\BibitemShut {NoStop}%
\bibitem [{\citenamefont {Chiao}\ \emph {et~al.}(1964)\citenamefont {Chiao}, \citenamefont {Townes},\ and\ \citenamefont {Stoicheff}}]{PhysRevLett.12.592}%
  \BibitemOpen
  \bibfield  {author} {\bibinfo {author} {\bibfnamefont {R.~Y.}\ \bibnamefont {Chiao}}, \bibinfo {author} {\bibfnamefont {C.~H.}\ \bibnamefont {Townes}},\ and\ \bibinfo {author} {\bibfnamefont {B.~P.}\ \bibnamefont {Stoicheff}},\ }\bibfield  {title} {\bibinfo {title} {Stimulated {B}rillouin scattering and coherent generation of intense hypersonic waves},\ }\href {https://doi.org/10.1103/PhysRevLett.12.592} {\bibfield  {journal} {\bibinfo  {journal} {Phys. Rev. Lett.}\ }\textbf {\bibinfo {volume} {12}},\ \bibinfo {pages} {592} (\bibinfo {year} {1964})}\BibitemShut {NoStop}%
\bibitem [{\citenamefont {Yang}\ \emph {et~al.}(2023)\citenamefont {Yang}, \citenamefont {Bevilacqua}, \citenamefont {Hambura}, \citenamefont {Neves}, \citenamefont {Gopalan}, \citenamefont {Watanabe}, \citenamefont {Govendir}, \citenamefont {Bernabeu}, \citenamefont {Ellenberg}, \citenamefont {Diz-Mu{\~n}oz} \emph {et~al.}}]{yang2023pulsed}%
  \BibitemOpen
  \bibfield  {author} {\bibinfo {author} {\bibfnamefont {F.}~\bibnamefont {Yang}}, \bibinfo {author} {\bibfnamefont {C.}~\bibnamefont {Bevilacqua}}, \bibinfo {author} {\bibfnamefont {S.}~\bibnamefont {Hambura}}, \bibinfo {author} {\bibfnamefont {A.}~\bibnamefont {Neves}}, \bibinfo {author} {\bibfnamefont {A.}~\bibnamefont {Gopalan}}, \bibinfo {author} {\bibfnamefont {K.}~\bibnamefont {Watanabe}}, \bibinfo {author} {\bibfnamefont {M.}~\bibnamefont {Govendir}}, \bibinfo {author} {\bibfnamefont {M.}~\bibnamefont {Bernabeu}}, \bibinfo {author} {\bibfnamefont {J.}~\bibnamefont {Ellenberg}}, \bibinfo {author} {\bibfnamefont {A.}~\bibnamefont {Diz-Mu{\~n}oz}}, \emph {et~al.},\ }\bibfield  {title} {\bibinfo {title} {Pulsed stimulated {B}rillouin microscopy enables high-sensitivity mechanical imaging of live and fragile biological specimens},\ }\href@noop {} {\bibfield  {journal} {\bibinfo  {journal} {Nature Methods}\ }\textbf {\bibinfo {volume} {20}},\ \bibinfo {pages} {1971} (\bibinfo {year} {2023})}\BibitemShut
  {NoStop}%
\bibitem [{\citenamefont {Merklein}\ \emph {et~al.}(2022)\citenamefont {Merklein}, \citenamefont {Kabakova}, \citenamefont {Zarifi},\ and\ \citenamefont {Eggleton}}]{merklein2022100}%
  \BibitemOpen
  \bibfield  {author} {\bibinfo {author} {\bibfnamefont {M.}~\bibnamefont {Merklein}}, \bibinfo {author} {\bibfnamefont {I.~V.}\ \bibnamefont {Kabakova}}, \bibinfo {author} {\bibfnamefont {A.}~\bibnamefont {Zarifi}},\ and\ \bibinfo {author} {\bibfnamefont {B.~J.}\ \bibnamefont {Eggleton}},\ }\bibfield  {title} {\bibinfo {title} {100 years of {B}rillouin scattering: Historical and future perspectives},\ }\href@noop {} {\bibfield  {journal} {\bibinfo  {journal} {Applied Physics Reviews}\ }\textbf {\bibinfo {volume} {9}} (\bibinfo {year} {2022})}\BibitemShut {NoStop}%
\bibitem [{\citenamefont {Remer}\ \emph {et~al.}(2020)\citenamefont {Remer}, \citenamefont {Shaashoua}, \citenamefont {Shemesh}, \citenamefont {Ben-Zvi},\ and\ \citenamefont {Bilenca}}]{remer2020high}%
  \BibitemOpen
  \bibfield  {author} {\bibinfo {author} {\bibfnamefont {I.}~\bibnamefont {Remer}}, \bibinfo {author} {\bibfnamefont {R.}~\bibnamefont {Shaashoua}}, \bibinfo {author} {\bibfnamefont {N.}~\bibnamefont {Shemesh}}, \bibinfo {author} {\bibfnamefont {A.}~\bibnamefont {Ben-Zvi}},\ and\ \bibinfo {author} {\bibfnamefont {A.}~\bibnamefont {Bilenca}},\ }\bibfield  {title} {\bibinfo {title} {High-sensitivity and high-specificity biomechanical imaging by stimulated {B}rillouin scattering microscopy},\ }\href@noop {} {\bibfield  {journal} {\bibinfo  {journal} {Nature Methods}\ }\textbf {\bibinfo {volume} {17}},\ \bibinfo {pages} {913} (\bibinfo {year} {2020})}\BibitemShut {NoStop}%
\bibitem [{\citenamefont {Ballmann}\ \emph {et~al.}(2017)\citenamefont {Ballmann}, \citenamefont {Meng}, \citenamefont {Traverso}, \citenamefont {Scully},\ and\ \citenamefont {Yakovlev}}]{ballmann2017impulsive}%
  \BibitemOpen
  \bibfield  {author} {\bibinfo {author} {\bibfnamefont {C.~W.}\ \bibnamefont {Ballmann}}, \bibinfo {author} {\bibfnamefont {Z.}~\bibnamefont {Meng}}, \bibinfo {author} {\bibfnamefont {A.~J.}\ \bibnamefont {Traverso}}, \bibinfo {author} {\bibfnamefont {M.~O.}\ \bibnamefont {Scully}},\ and\ \bibinfo {author} {\bibfnamefont {V.~V.}\ \bibnamefont {Yakovlev}},\ }\bibfield  {title} {\bibinfo {title} {Impulsive {B}rillouin microscopy},\ }\href@noop {} {\bibfield  {journal} {\bibinfo  {journal} {Optica}\ }\textbf {\bibinfo {volume} {4}},\ \bibinfo {pages} {124} (\bibinfo {year} {2017})}\BibitemShut {NoStop}%
\bibitem [{\citenamefont {Ballmann}\ \emph {et~al.}(2015)\citenamefont {Ballmann}, \citenamefont {Thompson}, \citenamefont {Traverso}, \citenamefont {Meng}, \citenamefont {Scully},\ and\ \citenamefont {Yakovlev}}]{ballmann2015stimulated}%
  \BibitemOpen
  \bibfield  {author} {\bibinfo {author} {\bibfnamefont {C.~W.}\ \bibnamefont {Ballmann}}, \bibinfo {author} {\bibfnamefont {J.~V.}\ \bibnamefont {Thompson}}, \bibinfo {author} {\bibfnamefont {A.~J.}\ \bibnamefont {Traverso}}, \bibinfo {author} {\bibfnamefont {Z.}~\bibnamefont {Meng}}, \bibinfo {author} {\bibfnamefont {M.~O.}\ \bibnamefont {Scully}},\ and\ \bibinfo {author} {\bibfnamefont {V.~V.}\ \bibnamefont {Yakovlev}},\ }\bibfield  {title} {\bibinfo {title} {Stimulated {B}rillouin scattering microscopic imaging},\ }\href@noop {} {\bibfield  {journal} {\bibinfo  {journal} {Scientific Reports}\ }\textbf {\bibinfo {volume} {5}},\ \bibinfo {pages} {1} (\bibinfo {year} {2015})}\BibitemShut {NoStop}%
\bibitem [{\citenamefont {Bevilacqua}\ \emph {et~al.}(2019)\citenamefont {Bevilacqua}, \citenamefont {S{\'a}nchez-Iranzo}, \citenamefont {Richter}, \citenamefont {Diz-Mu{\~n}oz},\ and\ \citenamefont {Prevedel}}]{bevilacqua2019imaging}%
  \BibitemOpen
  \bibfield  {author} {\bibinfo {author} {\bibfnamefont {C.}~\bibnamefont {Bevilacqua}}, \bibinfo {author} {\bibfnamefont {H.}~\bibnamefont {S{\'a}nchez-Iranzo}}, \bibinfo {author} {\bibfnamefont {D.}~\bibnamefont {Richter}}, \bibinfo {author} {\bibfnamefont {A.}~\bibnamefont {Diz-Mu{\~n}oz}},\ and\ \bibinfo {author} {\bibfnamefont {R.}~\bibnamefont {Prevedel}},\ }\bibfield  {title} {\bibinfo {title} {Imaging mechanical properties of sub-micron ecm in live zebrafish using {B}rillouin microscopy},\ }\href@noop {} {\bibfield  {journal} {\bibinfo  {journal} {Biomedical Optics Express}\ }\textbf {\bibinfo {volume} {10}},\ \bibinfo {pages} {1420} (\bibinfo {year} {2019})}\BibitemShut {NoStop}%
\bibitem [{\citenamefont {Raghunathan}\ \emph {et~al.}(2017)\citenamefont {Raghunathan}, \citenamefont {Zhang}, \citenamefont {Wu}, \citenamefont {Rippy}, \citenamefont {Singh}, \citenamefont {Larin},\ and\ \citenamefont {Scarcelli}}]{raghunathan2017evaluating}%
  \BibitemOpen
  \bibfield  {author} {\bibinfo {author} {\bibfnamefont {R.}~\bibnamefont {Raghunathan}}, \bibinfo {author} {\bibfnamefont {J.}~\bibnamefont {Zhang}}, \bibinfo {author} {\bibfnamefont {C.}~\bibnamefont {Wu}}, \bibinfo {author} {\bibfnamefont {J.}~\bibnamefont {Rippy}}, \bibinfo {author} {\bibfnamefont {M.}~\bibnamefont {Singh}}, \bibinfo {author} {\bibfnamefont {K.~V.}\ \bibnamefont {Larin}},\ and\ \bibinfo {author} {\bibfnamefont {G.}~\bibnamefont {Scarcelli}},\ }\bibfield  {title} {\bibinfo {title} {Evaluating biomechanical properties of murine embryos using {B}rillouin microscopy and optical coherence tomography},\ }\href@noop {} {\bibfield  {journal} {\bibinfo  {journal} {Journal of Biomedical Optics}\ }\textbf {\bibinfo {volume} {22}},\ \bibinfo {pages} {086013} (\bibinfo {year} {2017})}\BibitemShut {NoStop}%
\bibitem [{\citenamefont {Antonacci}\ and\ \citenamefont {Braakman}(2016)}]{antonacci2016biomechanics}%
  \BibitemOpen
  \bibfield  {author} {\bibinfo {author} {\bibfnamefont {G.}~\bibnamefont {Antonacci}}\ and\ \bibinfo {author} {\bibfnamefont {S.}~\bibnamefont {Braakman}},\ }\bibfield  {title} {\bibinfo {title} {Biomechanics of subcellular structures by non-invasive {B}rillouin microscopy},\ }\href@noop {} {\bibfield  {journal} {\bibinfo  {journal} {Scientific Reports}\ }\textbf {\bibinfo {volume} {6}},\ \bibinfo {pages} {37217} (\bibinfo {year} {2016})}\BibitemShut {NoStop}%
\bibitem [{\citenamefont {Palombo}\ \emph {et~al.}(2014{\natexlab{a}})\citenamefont {Palombo}, \citenamefont {Winlove}, \citenamefont {Edginton}, \citenamefont {Green}, \citenamefont {Stone}, \citenamefont {Caponi}, \citenamefont {Madami},\ and\ \citenamefont {Fioretto}}]{palombo2014biomechanics}%
  \BibitemOpen
  \bibfield  {author} {\bibinfo {author} {\bibfnamefont {F.}~\bibnamefont {Palombo}}, \bibinfo {author} {\bibfnamefont {C.~P.}\ \bibnamefont {Winlove}}, \bibinfo {author} {\bibfnamefont {R.~S.}\ \bibnamefont {Edginton}}, \bibinfo {author} {\bibfnamefont {E.}~\bibnamefont {Green}}, \bibinfo {author} {\bibfnamefont {N.}~\bibnamefont {Stone}}, \bibinfo {author} {\bibfnamefont {S.}~\bibnamefont {Caponi}}, \bibinfo {author} {\bibfnamefont {M.}~\bibnamefont {Madami}},\ and\ \bibinfo {author} {\bibfnamefont {D.}~\bibnamefont {Fioretto}},\ }\bibfield  {title} {\bibinfo {title} {Biomechanics of fibrous proteins of the extracellular matrix studied by {B}rillouin scattering},\ }\href@noop {} {\bibfield  {journal} {\bibinfo  {journal} {Journal of The Royal Society Interface}\ }\textbf {\bibinfo {volume} {11}},\ \bibinfo {pages} {20140739} (\bibinfo {year} {2014}{\natexlab{a}})}\BibitemShut {NoStop}%
\bibitem [{\citenamefont {Shi}\ \emph {et~al.}(2023)\citenamefont {Shi}, \citenamefont {Zhang},\ and\ \citenamefont {Zhang}}]{shi2023non}%
  \BibitemOpen
  \bibfield  {author} {\bibinfo {author} {\bibfnamefont {C.}~\bibnamefont {Shi}}, \bibinfo {author} {\bibfnamefont {H.}~\bibnamefont {Zhang}},\ and\ \bibinfo {author} {\bibfnamefont {J.}~\bibnamefont {Zhang}},\ }\bibfield  {title} {\bibinfo {title} {Non-contact and label-free biomechanical imaging: Stimulated {B}rillouin microscopy and beyond},\ }\href@noop {} {\bibfield  {journal} {\bibinfo  {journal} {Frontiers in Physics}\ }\textbf {\bibinfo {volume} {11}},\ \bibinfo {pages} {1175653} (\bibinfo {year} {2023})}\BibitemShut {NoStop}%
\bibitem [{\citenamefont {Koski}\ and\ \citenamefont {Yarger}(2005)}]{koski2005brillouin}%
  \BibitemOpen
  \bibfield  {author} {\bibinfo {author} {\bibfnamefont {K.}~\bibnamefont {Koski}}\ and\ \bibinfo {author} {\bibfnamefont {J.}~\bibnamefont {Yarger}},\ }\bibfield  {title} {\bibinfo {title} {Brillouin imaging},\ }\href@noop {} {\bibfield  {journal} {\bibinfo  {journal} {Applied Physics Letters}\ }\textbf {\bibinfo {volume} {87}} (\bibinfo {year} {2005})}\BibitemShut {NoStop}%
\bibitem [{\citenamefont {Shao}\ \emph {et~al.}(2018)\citenamefont {Shao}, \citenamefont {Seiler}, \citenamefont {Eltony}, \citenamefont {Ramier}, \citenamefont {Kwok}, \citenamefont {Scarcelli}, \citenamefont {Pineda~II},\ and\ \citenamefont {Yun}}]{shao2018effects}%
  \BibitemOpen
  \bibfield  {author} {\bibinfo {author} {\bibfnamefont {P.}~\bibnamefont {Shao}}, \bibinfo {author} {\bibfnamefont {T.~G.}\ \bibnamefont {Seiler}}, \bibinfo {author} {\bibfnamefont {A.~M.}\ \bibnamefont {Eltony}}, \bibinfo {author} {\bibfnamefont {A.}~\bibnamefont {Ramier}}, \bibinfo {author} {\bibfnamefont {S.~J.}\ \bibnamefont {Kwok}}, \bibinfo {author} {\bibfnamefont {G.}~\bibnamefont {Scarcelli}}, \bibinfo {author} {\bibfnamefont {R.}~\bibnamefont {Pineda~II}},\ and\ \bibinfo {author} {\bibfnamefont {S.-H.}\ \bibnamefont {Yun}},\ }\bibfield  {title} {\bibinfo {title} {Effects of corneal hydration on {B}rillouin microscopy in vivo},\ }\href@noop {} {\bibfield  {journal} {\bibinfo  {journal} {Investigative Ophthalmology \& Visual Science}\ }\textbf {\bibinfo {volume} {59}},\ \bibinfo {pages} {3020} (\bibinfo {year} {2018})}\BibitemShut {NoStop}%
\bibitem [{\citenamefont {Margueritat}\ \emph {et~al.}(2019)\citenamefont {Margueritat}, \citenamefont {Virgone-Carlotta}, \citenamefont {Monnier}, \citenamefont {Delano{\"e}-Ayari}, \citenamefont {Mertani}, \citenamefont {Berthelot}, \citenamefont {Martinet}, \citenamefont {Dagany}, \citenamefont {Rivi{\`e}re}, \citenamefont {Rieu} \emph {et~al.}}]{margueritat2019high}%
  \BibitemOpen
  \bibfield  {author} {\bibinfo {author} {\bibfnamefont {J.}~\bibnamefont {Margueritat}}, \bibinfo {author} {\bibfnamefont {A.}~\bibnamefont {Virgone-Carlotta}}, \bibinfo {author} {\bibfnamefont {S.}~\bibnamefont {Monnier}}, \bibinfo {author} {\bibfnamefont {H.}~\bibnamefont {Delano{\"e}-Ayari}}, \bibinfo {author} {\bibfnamefont {H.~C.}\ \bibnamefont {Mertani}}, \bibinfo {author} {\bibfnamefont {A.}~\bibnamefont {Berthelot}}, \bibinfo {author} {\bibfnamefont {Q.}~\bibnamefont {Martinet}}, \bibinfo {author} {\bibfnamefont {X.}~\bibnamefont {Dagany}}, \bibinfo {author} {\bibfnamefont {C.}~\bibnamefont {Rivi{\`e}re}}, \bibinfo {author} {\bibfnamefont {J.-P.}\ \bibnamefont {Rieu}}, \emph {et~al.},\ }\bibfield  {title} {\bibinfo {title} {High-frequency mechanical properties of tumors measured by {B}rillouin light scattering},\ }\href@noop {} {\bibfield  {journal} {\bibinfo  {journal} {Physical Review Letters}\ }\textbf {\bibinfo {volume} {122}},\ \bibinfo {pages} {018101} (\bibinfo {year} {2019})}\BibitemShut
  {NoStop}%
\bibitem [{\citenamefont {Conrad}\ \emph {et~al.}(2019)\citenamefont {Conrad}, \citenamefont {Gray}, \citenamefont {Stroka}, \citenamefont {Rizvi},\ and\ \citenamefont {Scarcelli}}]{conrad2019mechanical}%
  \BibitemOpen
  \bibfield  {author} {\bibinfo {author} {\bibfnamefont {C.}~\bibnamefont {Conrad}}, \bibinfo {author} {\bibfnamefont {K.}~\bibnamefont {Gray}}, \bibinfo {author} {\bibfnamefont {K.}~\bibnamefont {Stroka}}, \bibinfo {author} {\bibfnamefont {I.}~\bibnamefont {Rizvi}},\ and\ \bibinfo {author} {\bibfnamefont {G.}~\bibnamefont {Scarcelli}},\ }\bibfield  {title} {\bibinfo {title} {Mechanical characterization of 3d ovarian cancer nodules using {B}rillouin confocal microscopy},\ }\href@noop {} {\bibfield  {journal} {\bibinfo  {journal} {Cell Mol. Bioeng.}\ }\textbf {\bibinfo {volume} {12}},\ \bibinfo {pages} {215} (\bibinfo {year} {2019})}\BibitemShut {NoStop}%
\bibitem [{\citenamefont {Palombo}\ \emph {et~al.}(2014{\natexlab{b}})\citenamefont {Palombo}, \citenamefont {Madami}, \citenamefont {Stone},\ and\ \citenamefont {Fioretto}}]{palombo2014mechanical}%
  \BibitemOpen
  \bibfield  {author} {\bibinfo {author} {\bibfnamefont {F.}~\bibnamefont {Palombo}}, \bibinfo {author} {\bibfnamefont {M.}~\bibnamefont {Madami}}, \bibinfo {author} {\bibfnamefont {N.}~\bibnamefont {Stone}},\ and\ \bibinfo {author} {\bibfnamefont {D.}~\bibnamefont {Fioretto}},\ }\bibfield  {title} {\bibinfo {title} {Mechanical mapping with chemical specificity by confocal {B}rillouin and {R}aman microscopy},\ }\href@noop {} {\bibfield  {journal} {\bibinfo  {journal} {Analyst}\ }\textbf {\bibinfo {volume} {139}},\ \bibinfo {pages} {729} (\bibinfo {year} {2014}{\natexlab{b}})}\BibitemShut {NoStop}%
\bibitem [{\citenamefont {Zhang}\ \emph {et~al.}(2016)\citenamefont {Zhang}, \citenamefont {Fiore}, \citenamefont {Yun}, \citenamefont {Kim},\ and\ \citenamefont {Scarcelli}}]{zhang2016line}%
  \BibitemOpen
  \bibfield  {author} {\bibinfo {author} {\bibfnamefont {J.}~\bibnamefont {Zhang}}, \bibinfo {author} {\bibfnamefont {A.}~\bibnamefont {Fiore}}, \bibinfo {author} {\bibfnamefont {S.-H.}\ \bibnamefont {Yun}}, \bibinfo {author} {\bibfnamefont {H.}~\bibnamefont {Kim}},\ and\ \bibinfo {author} {\bibfnamefont {G.}~\bibnamefont {Scarcelli}},\ }\bibfield  {title} {\bibinfo {title} {Line-scanning {B}rillouin microscopy for rapid non-invasive mechanical imaging},\ }\href@noop {} {\bibfield  {journal} {\bibinfo  {journal} {Scientific Reports}\ }\textbf {\bibinfo {volume} {6}},\ \bibinfo {pages} {35398} (\bibinfo {year} {2016})}\BibitemShut {NoStop}%
\bibitem [{\citenamefont {Garmire}(2017)}]{garmire2017perspectives}%
  \BibitemOpen
  \bibfield  {author} {\bibinfo {author} {\bibfnamefont {E.}~\bibnamefont {Garmire}},\ }\bibfield  {title} {\bibinfo {title} {Perspectives on stimulated {B}rillouin scattering},\ }\href@noop {} {\bibfield  {journal} {\bibinfo  {journal} {New Journal of Physics}\ }\textbf {\bibinfo {volume} {19}},\ \bibinfo {pages} {011003} (\bibinfo {year} {2017})}\BibitemShut {NoStop}%
\bibitem [{\citenamefont {Casacio}\ \emph {et~al.}(2021)\citenamefont {Casacio}, \citenamefont {Madsen}, \citenamefont {Terrasson}, \citenamefont {Waleed}, \citenamefont {Barnscheidt}, \citenamefont {Hage}, \citenamefont {Taylor},\ and\ \citenamefont {Bowen}}]{casacio2021quantum}%
  \BibitemOpen
  \bibfield  {author} {\bibinfo {author} {\bibfnamefont {C.~A.}\ \bibnamefont {Casacio}}, \bibinfo {author} {\bibfnamefont {L.~S.}\ \bibnamefont {Madsen}}, \bibinfo {author} {\bibfnamefont {A.}~\bibnamefont {Terrasson}}, \bibinfo {author} {\bibfnamefont {M.}~\bibnamefont {Waleed}}, \bibinfo {author} {\bibfnamefont {K.}~\bibnamefont {Barnscheidt}}, \bibinfo {author} {\bibfnamefont {B.}~\bibnamefont {Hage}}, \bibinfo {author} {\bibfnamefont {M.~A.}\ \bibnamefont {Taylor}},\ and\ \bibinfo {author} {\bibfnamefont {W.~P.}\ \bibnamefont {Bowen}},\ }\bibfield  {title} {\bibinfo {title} {Quantum-enhanced nonlinear microscopy},\ }\href@noop {} {\bibfield  {journal} {\bibinfo  {journal} {Nature}\ }\textbf {\bibinfo {volume} {594}},\ \bibinfo {pages} {201} (\bibinfo {year} {2021})}\BibitemShut {NoStop}%
\bibitem [{\citenamefont {Taylor}\ \emph {et~al.}(2013)\citenamefont {Taylor}, \citenamefont {Janousek}, \citenamefont {Daria}, \citenamefont {Knittel}, \citenamefont {Hage}, \citenamefont {Bachor},\ and\ \citenamefont {Bowen}}]{taylor2013biological}%
  \BibitemOpen
  \bibfield  {author} {\bibinfo {author} {\bibfnamefont {M.~A.}\ \bibnamefont {Taylor}}, \bibinfo {author} {\bibfnamefont {J.}~\bibnamefont {Janousek}}, \bibinfo {author} {\bibfnamefont {V.}~\bibnamefont {Daria}}, \bibinfo {author} {\bibfnamefont {J.}~\bibnamefont {Knittel}}, \bibinfo {author} {\bibfnamefont {B.}~\bibnamefont {Hage}}, \bibinfo {author} {\bibfnamefont {H.-A.}\ \bibnamefont {Bachor}},\ and\ \bibinfo {author} {\bibfnamefont {W.~P.}\ \bibnamefont {Bowen}},\ }\bibfield  {title} {\bibinfo {title} {Biological measurement beyond the quantum limit},\ }\href@noop {} {\bibfield  {journal} {\bibinfo  {journal} {Nature Photonics}\ }\textbf {\bibinfo {volume} {7}},\ \bibinfo {pages} {229} (\bibinfo {year} {2013})}\BibitemShut {NoStop}%
\bibitem [{\citenamefont {Taylor}\ and\ \citenamefont {Bowen}(2016)}]{taylor2016quantum}%
  \BibitemOpen
  \bibfield  {author} {\bibinfo {author} {\bibfnamefont {M.~A.}\ \bibnamefont {Taylor}}\ and\ \bibinfo {author} {\bibfnamefont {W.~P.}\ \bibnamefont {Bowen}},\ }\bibfield  {title} {\bibinfo {title} {Quantum metrology and its application in biology},\ }\href@noop {} {\bibfield  {journal} {\bibinfo  {journal} {Physics Reports}\ }\textbf {\bibinfo {volume} {615}},\ \bibinfo {pages} {1} (\bibinfo {year} {2016})}\BibitemShut {NoStop}%
\bibitem [{\citenamefont {de~Andrade}\ \emph {et~al.}(2020)\citenamefont {de~Andrade}, \citenamefont {Kerdoncuff}, \citenamefont {Berg-S{\o}rensen}, \citenamefont {Gehring}, \citenamefont {Lassen},\ and\ \citenamefont {Andersen}}]{de2020quantum}%
  \BibitemOpen
  \bibfield  {author} {\bibinfo {author} {\bibfnamefont {R.~B.}\ \bibnamefont {de~Andrade}}, \bibinfo {author} {\bibfnamefont {H.}~\bibnamefont {Kerdoncuff}}, \bibinfo {author} {\bibfnamefont {K.}~\bibnamefont {Berg-S{\o}rensen}}, \bibinfo {author} {\bibfnamefont {T.}~\bibnamefont {Gehring}}, \bibinfo {author} {\bibfnamefont {M.}~\bibnamefont {Lassen}},\ and\ \bibinfo {author} {\bibfnamefont {U.~L.}\ \bibnamefont {Andersen}},\ }\bibfield  {title} {\bibinfo {title} {Quantum-enhanced continuous-wave stimulated {R}aman scattering spectroscopy},\ }\href@noop {} {\bibfield  {journal} {\bibinfo  {journal} {Optica}\ }\textbf {\bibinfo {volume} {7}},\ \bibinfo {pages} {470} (\bibinfo {year} {2020})}\BibitemShut {NoStop}%
\bibitem [{\citenamefont {Varnavski}\ \emph {et~al.}(2022)\citenamefont {Varnavski}, \citenamefont {Gunthardt}, \citenamefont {Rehman}, \citenamefont {Luker},\ and\ \citenamefont {Goodson~III}}]{varnavski2022quantum}%
  \BibitemOpen
  \bibfield  {author} {\bibinfo {author} {\bibfnamefont {O.}~\bibnamefont {Varnavski}}, \bibinfo {author} {\bibfnamefont {C.}~\bibnamefont {Gunthardt}}, \bibinfo {author} {\bibfnamefont {A.}~\bibnamefont {Rehman}}, \bibinfo {author} {\bibfnamefont {G.~D.}\ \bibnamefont {Luker}},\ and\ \bibinfo {author} {\bibfnamefont {T.}~\bibnamefont {Goodson~III}},\ }\bibfield  {title} {\bibinfo {title} {Quantum light-enhanced two-photon imaging of breast cancer cells},\ }\href@noop {} {\bibfield  {journal} {\bibinfo  {journal} {The Journal of Physical Chemistry Letters}\ }\textbf {\bibinfo {volume} {13}},\ \bibinfo {pages} {2772} (\bibinfo {year} {2022})}\BibitemShut {NoStop}%
\bibitem [{\citenamefont {Schlawin}(2017)}]{schlawin2017entangled}%
  \BibitemOpen
  \bibfield  {author} {\bibinfo {author} {\bibfnamefont {F.}~\bibnamefont {Schlawin}},\ }\bibfield  {title} {\bibinfo {title} {Entangled photon spectroscopy},\ }\href@noop {} {\bibfield  {journal} {\bibinfo  {journal} {Journal of Physics B: Atomic, Molecular and Optical Physics}\ }\textbf {\bibinfo {volume} {50}},\ \bibinfo {pages} {203001} (\bibinfo {year} {2017})}\BibitemShut {NoStop}%
\bibitem [{\citenamefont {Burdick}\ \emph {et~al.}(2021)\citenamefont {Burdick}, \citenamefont {Schatz},\ and\ \citenamefont {Goodson~III}}]{burdick2021enhancing}%
  \BibitemOpen
  \bibfield  {author} {\bibinfo {author} {\bibfnamefont {R.~K.}\ \bibnamefont {Burdick}}, \bibinfo {author} {\bibfnamefont {G.~C.}\ \bibnamefont {Schatz}},\ and\ \bibinfo {author} {\bibfnamefont {T.}~\bibnamefont {Goodson~III}},\ }\bibfield  {title} {\bibinfo {title} {Enhancing entangled two-photon absorption for picosecond quantum spectroscopy},\ }\href@noop {} {\bibfield  {journal} {\bibinfo  {journal} {Journal of the American Chemical Society}\ }\textbf {\bibinfo {volume} {143}},\ \bibinfo {pages} {16930} (\bibinfo {year} {2021})}\BibitemShut {NoStop}%
\bibitem [{\citenamefont {Varnavski}\ and\ \citenamefont {Goodson~III}(2020)}]{varnavski2020two}%
  \BibitemOpen
  \bibfield  {author} {\bibinfo {author} {\bibfnamefont {O.}~\bibnamefont {Varnavski}}\ and\ \bibinfo {author} {\bibfnamefont {T.}~\bibnamefont {Goodson~III}},\ }\bibfield  {title} {\bibinfo {title} {Two-photon fluorescence microscopy at extremely low excitation intensity: The power of quantum correlations},\ }\href@noop {} {\bibfield  {journal} {\bibinfo  {journal} {Journal of the American Chemical Society}\ }\textbf {\bibinfo {volume} {142}},\ \bibinfo {pages} {12966} (\bibinfo {year} {2020})}\BibitemShut {NoStop}%
\bibitem [{\citenamefont {Villabona-Monsalve}\ \emph {et~al.}(2020)\citenamefont {Villabona-Monsalve}, \citenamefont {Burdick},\ and\ \citenamefont {Goodson~III}}]{villabona2020measurements}%
  \BibitemOpen
  \bibfield  {author} {\bibinfo {author} {\bibfnamefont {J.~P.}\ \bibnamefont {Villabona-Monsalve}}, \bibinfo {author} {\bibfnamefont {R.~K.}\ \bibnamefont {Burdick}},\ and\ \bibinfo {author} {\bibfnamefont {T.}~\bibnamefont {Goodson~III}},\ }\bibfield  {title} {\bibinfo {title} {Measurements of entangled two-photon absorption in organic molecules with cw-pumped type-i spontaneous parametric down-conversion},\ }\href@noop {} {\bibfield  {journal} {\bibinfo  {journal} {The Journal of Physical Chemistry C}\ }\textbf {\bibinfo {volume} {124}},\ \bibinfo {pages} {24526} (\bibinfo {year} {2020})}\BibitemShut {NoStop}%
\bibitem [{\citenamefont {Varnavski}\ \emph {et~al.}(2017)\citenamefont {Varnavski}, \citenamefont {Pinsky},\ and\ \citenamefont {Goodson~III}}]{varnavski2017entangled}%
  \BibitemOpen
  \bibfield  {author} {\bibinfo {author} {\bibfnamefont {O.}~\bibnamefont {Varnavski}}, \bibinfo {author} {\bibfnamefont {B.}~\bibnamefont {Pinsky}},\ and\ \bibinfo {author} {\bibfnamefont {T.}~\bibnamefont {Goodson~III}},\ }\bibfield  {title} {\bibinfo {title} {Entangled photon excited fluorescence in organic materials: an ultrafast coincidence detector},\ }\href@noop {} {\bibfield  {journal} {\bibinfo  {journal} {The journal of physical chemistry letters}\ }\textbf {\bibinfo {volume} {8}},\ \bibinfo {pages} {388} (\bibinfo {year} {2017})}\BibitemShut {NoStop}%
\bibitem [{\citenamefont {Schlawin}\ \emph {et~al.}(2018)\citenamefont {Schlawin}, \citenamefont {Dorfman},\ and\ \citenamefont {Mukamel}}]{schlawin2018entangled}%
  \BibitemOpen
  \bibfield  {author} {\bibinfo {author} {\bibfnamefont {F.}~\bibnamefont {Schlawin}}, \bibinfo {author} {\bibfnamefont {K.~E.}\ \bibnamefont {Dorfman}},\ and\ \bibinfo {author} {\bibfnamefont {S.}~\bibnamefont {Mukamel}},\ }\bibfield  {title} {\bibinfo {title} {Entangled two-photon absorption spectroscopy},\ }\href@noop {} {\bibfield  {journal} {\bibinfo  {journal} {Accounts of chemical research}\ }\textbf {\bibinfo {volume} {51}},\ \bibinfo {pages} {2207} (\bibinfo {year} {2018})}\BibitemShut {NoStop}%
\bibitem [{\citenamefont {Dorfman}\ \emph {et~al.}(2016)\citenamefont {Dorfman}, \citenamefont {Schlawin},\ and\ \citenamefont {Mukamel}}]{dorfman2016nonlinear}%
  \BibitemOpen
  \bibfield  {author} {\bibinfo {author} {\bibfnamefont {K.~E.}\ \bibnamefont {Dorfman}}, \bibinfo {author} {\bibfnamefont {F.}~\bibnamefont {Schlawin}},\ and\ \bibinfo {author} {\bibfnamefont {S.}~\bibnamefont {Mukamel}},\ }\bibfield  {title} {\bibinfo {title} {Nonlinear optical signals and spectroscopy with quantum light},\ }\href@noop {} {\bibfield  {journal} {\bibinfo  {journal} {Reviews of Modern Physics}\ }\textbf {\bibinfo {volume} {88}},\ \bibinfo {pages} {045008} (\bibinfo {year} {2016})}\BibitemShut {NoStop}%
\bibitem [{\citenamefont {Varnavski}\ \emph {et~al.}(2023)\citenamefont {Varnavski}, \citenamefont {Giri}, \citenamefont {Chiang}, \citenamefont {Zeman~IV}, \citenamefont {Schatz},\ and\ \citenamefont {Goodson~III}}]{varnavski2023colors}%
  \BibitemOpen
  \bibfield  {author} {\bibinfo {author} {\bibfnamefont {O.}~\bibnamefont {Varnavski}}, \bibinfo {author} {\bibfnamefont {S.~K.}\ \bibnamefont {Giri}}, \bibinfo {author} {\bibfnamefont {T.-M.}\ \bibnamefont {Chiang}}, \bibinfo {author} {\bibfnamefont {C.~J.}\ \bibnamefont {Zeman~IV}}, \bibinfo {author} {\bibfnamefont {G.~C.}\ \bibnamefont {Schatz}},\ and\ \bibinfo {author} {\bibfnamefont {T.}~\bibnamefont {Goodson~III}},\ }\bibfield  {title} {\bibinfo {title} {Colors of entangled two-photon absorption},\ }\href@noop {} {\bibfield  {journal} {\bibinfo  {journal} {Proceedings of the National Academy of Sciences}\ }\textbf {\bibinfo {volume} {120}},\ \bibinfo {pages} {e2307719120} (\bibinfo {year} {2023})}\BibitemShut {NoStop}%
\bibitem [{\citenamefont {Li}\ \emph {et~al.}(2022)\citenamefont {Li}, \citenamefont {Li}, \citenamefont {Liu}, \citenamefont {Yakovlev},\ and\ \citenamefont {Agarwal}}]{li2022quantum}%
  \BibitemOpen
  \bibfield  {author} {\bibinfo {author} {\bibfnamefont {T.}~\bibnamefont {Li}}, \bibinfo {author} {\bibfnamefont {F.}~\bibnamefont {Li}}, \bibinfo {author} {\bibfnamefont {X.}~\bibnamefont {Liu}}, \bibinfo {author} {\bibfnamefont {V.~V.}\ \bibnamefont {Yakovlev}},\ and\ \bibinfo {author} {\bibfnamefont {G.~S.}\ \bibnamefont {Agarwal}},\ }\bibfield  {title} {\bibinfo {title} {Quantum-enhanced stimulated {B}rillouin scattering spectroscopy and imaging},\ }\href@noop {} {\bibfield  {journal} {\bibinfo  {journal} {Optica}\ }\textbf {\bibinfo {volume} {9}},\ \bibinfo {pages} {959} (\bibinfo {year} {2022})}\BibitemShut {NoStop}%
\bibitem [{\citenamefont {Scully}\ and\ \citenamefont {Zubairy}(1997)}]{scully1997quantum}%
  \BibitemOpen
  \bibfield  {author} {\bibinfo {author} {\bibfnamefont {M.~O.}\ \bibnamefont {Scully}}\ and\ \bibinfo {author} {\bibfnamefont {M.~S.}\ \bibnamefont {Zubairy}},\ }\href@noop {} {\emph {\bibinfo {title} {Quantum Optics}}}\ (\bibinfo  {publisher} {Cambridge University Press},\ \bibinfo {year} {1997})\BibitemShut {NoStop}%
\bibitem [{\citenamefont {Dorfman}\ \emph {et~al.}(2021)\citenamefont {Dorfman}, \citenamefont {Liu}, \citenamefont {Lou}, \citenamefont {Wei}, \citenamefont {Jing}, \citenamefont {Schlawin},\ and\ \citenamefont {Mukamel}}]{dorfman2021multidimensional}%
  \BibitemOpen
  \bibfield  {author} {\bibinfo {author} {\bibfnamefont {K.}~\bibnamefont {Dorfman}}, \bibinfo {author} {\bibfnamefont {S.}~\bibnamefont {Liu}}, \bibinfo {author} {\bibfnamefont {Y.}~\bibnamefont {Lou}}, \bibinfo {author} {\bibfnamefont {T.}~\bibnamefont {Wei}}, \bibinfo {author} {\bibfnamefont {J.}~\bibnamefont {Jing}}, \bibinfo {author} {\bibfnamefont {F.}~\bibnamefont {Schlawin}},\ and\ \bibinfo {author} {\bibfnamefont {S.}~\bibnamefont {Mukamel}},\ }\bibfield  {title} {\bibinfo {title} {Multidimensional four-wave mixing signals detected by quantum squeezed light},\ }\href@noop {} {\bibfield  {journal} {\bibinfo  {journal} {Proceedings of the National Academy of Sciences}\ }\textbf {\bibinfo {volume} {118}} (\bibinfo {year} {2021})}\BibitemShut {NoStop}%
\bibitem [{\citenamefont {Li}\ \emph {et~al.}(2021{\natexlab{a}})\citenamefont {Li}, \citenamefont {Li}, \citenamefont {Scully},\ and\ \citenamefont {Agarwal}}]{PhysRevApplied.15.044030}%
  \BibitemOpen
  \bibfield  {author} {\bibinfo {author} {\bibfnamefont {F.}~\bibnamefont {Li}}, \bibinfo {author} {\bibfnamefont {T.}~\bibnamefont {Li}}, \bibinfo {author} {\bibfnamefont {M.~O.}\ \bibnamefont {Scully}},\ and\ \bibinfo {author} {\bibfnamefont {G.~S.}\ \bibnamefont {Agarwal}},\ }\bibfield  {title} {\bibinfo {title} {Quantum advantage with seeded squeezed light for absorption measurement},\ }\href {https://doi.org/10.1103/PhysRevApplied.15.044030} {\bibfield  {journal} {\bibinfo  {journal} {Phys. Rev. Applied}\ }\textbf {\bibinfo {volume} {15}},\ \bibinfo {pages} {044030} (\bibinfo {year} {2021}{\natexlab{a}})}\BibitemShut {NoStop}%
\bibitem [{\citenamefont {Prajapati}\ \emph {et~al.}(2021)\citenamefont {Prajapati}, \citenamefont {Niu},\ and\ \citenamefont {Novikova}}]{prajapati2021quantum}%
  \BibitemOpen
  \bibfield  {author} {\bibinfo {author} {\bibfnamefont {N.}~\bibnamefont {Prajapati}}, \bibinfo {author} {\bibfnamefont {Z.}~\bibnamefont {Niu}},\ and\ \bibinfo {author} {\bibfnamefont {I.}~\bibnamefont {Novikova}},\ }\bibfield  {title} {\bibinfo {title} {Quantum-enhanced two-photon spectroscopy using two-mode squeezed light},\ }\href@noop {} {\bibfield  {journal} {\bibinfo  {journal} {Optics Letters}\ }\textbf {\bibinfo {volume} {46}},\ \bibinfo {pages} {1800} (\bibinfo {year} {2021})}\BibitemShut {NoStop}%
\bibitem [{\citenamefont {Dowran}\ \emph {et~al.}(2018)\citenamefont {Dowran}, \citenamefont {Kumar}, \citenamefont {Lawrie}, \citenamefont {Pooser},\ and\ \citenamefont {Marino}}]{Dowran:18}%
  \BibitemOpen
  \bibfield  {author} {\bibinfo {author} {\bibfnamefont {M.}~\bibnamefont {Dowran}}, \bibinfo {author} {\bibfnamefont {A.}~\bibnamefont {Kumar}}, \bibinfo {author} {\bibfnamefont {B.~J.}\ \bibnamefont {Lawrie}}, \bibinfo {author} {\bibfnamefont {R.~C.}\ \bibnamefont {Pooser}},\ and\ \bibinfo {author} {\bibfnamefont {A.~M.}\ \bibnamefont {Marino}},\ }\bibfield  {title} {\bibinfo {title} {Quantum-enhanced plasmonic sensing},\ }\href {https://doi.org/10.1364/OPTICA.5.000628} {\bibfield  {journal} {\bibinfo  {journal} {Optica}\ }\textbf {\bibinfo {volume} {5}},\ \bibinfo {pages} {628} (\bibinfo {year} {2018})}\BibitemShut {NoStop}%
\bibitem [{\citenamefont {Anderson}\ \emph {et~al.}(2017)\citenamefont {Anderson}, \citenamefont {Gupta}, \citenamefont {Schmittberger}, \citenamefont {Horrom}, \citenamefont {Hermann-Avigliano}, \citenamefont {Jones},\ and\ \citenamefont {Lett}}]{Anderson:17}%
  \BibitemOpen
  \bibfield  {author} {\bibinfo {author} {\bibfnamefont {B.~E.}\ \bibnamefont {Anderson}}, \bibinfo {author} {\bibfnamefont {P.}~\bibnamefont {Gupta}}, \bibinfo {author} {\bibfnamefont {B.~L.}\ \bibnamefont {Schmittberger}}, \bibinfo {author} {\bibfnamefont {T.}~\bibnamefont {Horrom}}, \bibinfo {author} {\bibfnamefont {C.}~\bibnamefont {Hermann-Avigliano}}, \bibinfo {author} {\bibfnamefont {K.~M.}\ \bibnamefont {Jones}},\ and\ \bibinfo {author} {\bibfnamefont {P.~D.}\ \bibnamefont {Lett}},\ }\bibfield  {title} {\bibinfo {title} {Phase sensing beyond the standard quantum limit with a variation on the {SU}(1,1) interferometer},\ }\href {https://doi.org/10.1364/OPTICA.4.000752} {\bibfield  {journal} {\bibinfo  {journal} {Optica}\ }\textbf {\bibinfo {volume} {4}},\ \bibinfo {pages} {752} (\bibinfo {year} {2017})}\BibitemShut {NoStop}%
\bibitem [{\citenamefont {Pooser}\ and\ \citenamefont {Lawrie}(2015)}]{Pooser:15}%
  \BibitemOpen
  \bibfield  {author} {\bibinfo {author} {\bibfnamefont {R.~C.}\ \bibnamefont {Pooser}}\ and\ \bibinfo {author} {\bibfnamefont {B.}~\bibnamefont {Lawrie}},\ }\bibfield  {title} {\bibinfo {title} {Ultrasensitive measurement of microcantilever displacement below the shot-noise limit},\ }\href {https://doi.org/10.1364/OPTICA.2.000393} {\bibfield  {journal} {\bibinfo  {journal} {Optica}\ }\textbf {\bibinfo {volume} {2}},\ \bibinfo {pages} {393} (\bibinfo {year} {2015})}\BibitemShut {NoStop}%
\bibitem [{\citenamefont {Li}\ \emph {et~al.}(2021{\natexlab{b}})\citenamefont {Li}, \citenamefont {Li},\ and\ \citenamefont {Agarwal}}]{PhysRevResearch.3.033095}%
  \BibitemOpen
  \bibfield  {author} {\bibinfo {author} {\bibfnamefont {F.}~\bibnamefont {Li}}, \bibinfo {author} {\bibfnamefont {T.}~\bibnamefont {Li}},\ and\ \bibinfo {author} {\bibfnamefont {G.~S.}\ \bibnamefont {Agarwal}},\ }\bibfield  {title} {\bibinfo {title} {Experimental study of decoherence of the two-mode squeezed vacuum state via second harmonic generation},\ }\href {https://doi.org/10.1103/PhysRevResearch.3.033095} {\bibfield  {journal} {\bibinfo  {journal} {Phys. Rev. Research}\ }\textbf {\bibinfo {volume} {3}},\ \bibinfo {pages} {033095} (\bibinfo {year} {2021}{\natexlab{b}})}\BibitemShut {NoStop}%
\bibitem [{\citenamefont {Clark}\ \emph {et~al.}(2014)\citenamefont {Clark}, \citenamefont {Glasser}, \citenamefont {Glorieux}, \citenamefont {Vogl}, \citenamefont {Li}, \citenamefont {Jones},\ and\ \citenamefont {Lett}}]{Clark:2014vf}%
  \BibitemOpen
  \bibfield  {author} {\bibinfo {author} {\bibfnamefont {J.~B.}\ \bibnamefont {Clark}}, \bibinfo {author} {\bibfnamefont {R.~T.}\ \bibnamefont {Glasser}}, \bibinfo {author} {\bibfnamefont {Q.}~\bibnamefont {Glorieux}}, \bibinfo {author} {\bibfnamefont {U.}~\bibnamefont {Vogl}}, \bibinfo {author} {\bibfnamefont {T.}~\bibnamefont {Li}}, \bibinfo {author} {\bibfnamefont {K.~M.}\ \bibnamefont {Jones}},\ and\ \bibinfo {author} {\bibfnamefont {P.~D.}\ \bibnamefont {Lett}},\ }\bibfield  {title} {\bibinfo {title} {Quantum mutual information of an entangled state propagating through a fast-light medium},\ }\href@noop {} {\bibfield  {journal} {\bibinfo  {journal} {Nature Photonics}\ }\textbf {\bibinfo {volume} {8}},\ \bibinfo {pages} {515} (\bibinfo {year} {2014})}\BibitemShut {NoStop}%
\bibitem [{\citenamefont {Glasser}\ \emph {et~al.}(2012)\citenamefont {Glasser}, \citenamefont {Vogl},\ and\ \citenamefont {Lett}}]{Glasser2012a}%
  \BibitemOpen
  \bibfield  {author} {\bibinfo {author} {\bibfnamefont {R.~T.}\ \bibnamefont {Glasser}}, \bibinfo {author} {\bibfnamefont {U.}~\bibnamefont {Vogl}},\ and\ \bibinfo {author} {\bibfnamefont {P.~D.}\ \bibnamefont {Lett}},\ }\bibfield  {title} {\bibinfo {title} {{Stimulated Generation of Superluminal Light Pulses via Four-Wave Mixing}},\ }\href {https://doi.org/10.1103/PhysRevLett.108.173902} {\bibfield  {journal} {\bibinfo  {journal} {Physical Review Letters}\ }\textbf {\bibinfo {volume} {108}},\ \bibinfo {pages} {173902} (\bibinfo {year} {2012})}\BibitemShut {NoStop}%
\bibitem [{\citenamefont {Boyd}(2008)}]{NonlinearOptics}%
  \BibitemOpen
  \bibfield  {author} {\bibinfo {author} {\bibfnamefont {R.~W.}\ \bibnamefont {Boyd}},\ }\href@noop {} {\emph {\bibinfo {title} {Nonlinear Optics}}}\ (\bibinfo  {publisher} {Academic Press, Burlington, MA},\ \bibinfo {year} {2008})\BibitemShut {NoStop}%
\bibitem [{\citenamefont {Li}\ \emph {et~al.}(2020)\citenamefont {Li}, \citenamefont {Li}, \citenamefont {Altuzarra}, \citenamefont {Classen},\ and\ \citenamefont {Agarwal}}]{doi:10.1063/5.0010909}%
  \BibitemOpen
  \bibfield  {author} {\bibinfo {author} {\bibfnamefont {T.}~\bibnamefont {Li}}, \bibinfo {author} {\bibfnamefont {F.}~\bibnamefont {Li}}, \bibinfo {author} {\bibfnamefont {C.}~\bibnamefont {Altuzarra}}, \bibinfo {author} {\bibfnamefont {A.}~\bibnamefont {Classen}},\ and\ \bibinfo {author} {\bibfnamefont {G.~S.}\ \bibnamefont {Agarwal}},\ }\bibfield  {title} {\bibinfo {title} {Squeezed light induced two-photon absorption fluorescence of fluorescein biomarkers},\ }\href {https://doi.org/10.1063/5.0010909} {\bibfield  {journal} {\bibinfo  {journal} {Applied Physics Letters}\ }\textbf {\bibinfo {volume} {116}},\ \bibinfo {pages} {254001} (\bibinfo {year} {2020})}\BibitemShut {NoStop}%
\bibitem [{\citenamefont {Antonacci}\ \emph {et~al.}(2015)\citenamefont {Antonacci}, \citenamefont {Pedrigi}, \citenamefont {Kondiboyina}, \citenamefont {Mehta}, \citenamefont {De~Silva}, \citenamefont {Paterson}, \citenamefont {Krams},\ and\ \citenamefont {T{\"o}r{\"o}k}}]{antonacci2015quantification}%
  \BibitemOpen
  \bibfield  {author} {\bibinfo {author} {\bibfnamefont {G.}~\bibnamefont {Antonacci}}, \bibinfo {author} {\bibfnamefont {R.~M.}\ \bibnamefont {Pedrigi}}, \bibinfo {author} {\bibfnamefont {A.}~\bibnamefont {Kondiboyina}}, \bibinfo {author} {\bibfnamefont {V.~V.}\ \bibnamefont {Mehta}}, \bibinfo {author} {\bibfnamefont {R.}~\bibnamefont {De~Silva}}, \bibinfo {author} {\bibfnamefont {C.}~\bibnamefont {Paterson}}, \bibinfo {author} {\bibfnamefont {R.}~\bibnamefont {Krams}},\ and\ \bibinfo {author} {\bibfnamefont {P.}~\bibnamefont {T{\"o}r{\"o}k}},\ }\bibfield  {title} {\bibinfo {title} {Quantification of plaque stiffness by {B}rillouin microscopy in experimental thin cap fibroatheroma},\ }\href@noop {} {\bibfield  {journal} {\bibinfo  {journal} {Journal of the Royal Society Interface}\ }\textbf {\bibinfo {volume} {12}},\ \bibinfo {pages} {20150843} (\bibinfo {year} {2015})}\BibitemShut {NoStop}%
\bibitem [{\citenamefont {Park}\ \emph {et~al.}(2024)\citenamefont {Park}, \citenamefont {Stokowski}, \citenamefont {Ansari}, \citenamefont {Gyger}, \citenamefont {Multani}, \citenamefont {Celik}, \citenamefont {Hwang}, \citenamefont {Dean}, \citenamefont {Mayor}, \citenamefont {McKenna} \emph {et~al.}}]{park2024single}%
  \BibitemOpen
  \bibfield  {author} {\bibinfo {author} {\bibfnamefont {T.}~\bibnamefont {Park}}, \bibinfo {author} {\bibfnamefont {H.}~\bibnamefont {Stokowski}}, \bibinfo {author} {\bibfnamefont {V.}~\bibnamefont {Ansari}}, \bibinfo {author} {\bibfnamefont {S.}~\bibnamefont {Gyger}}, \bibinfo {author} {\bibfnamefont {K.~K.}\ \bibnamefont {Multani}}, \bibinfo {author} {\bibfnamefont {O.~T.}\ \bibnamefont {Celik}}, \bibinfo {author} {\bibfnamefont {A.~Y.}\ \bibnamefont {Hwang}}, \bibinfo {author} {\bibfnamefont {D.~J.}\ \bibnamefont {Dean}}, \bibinfo {author} {\bibfnamefont {F.}~\bibnamefont {Mayor}}, \bibinfo {author} {\bibfnamefont {T.~P.}\ \bibnamefont {McKenna}}, \emph {et~al.},\ }\bibfield  {title} {\bibinfo {title} {Single-mode squeezed-light generation and tomography with an integrated optical parametric oscillator},\ }\href@noop {} {\bibfield  {journal} {\bibinfo  {journal} {Science Advances}\ }\textbf {\bibinfo {volume} {10}},\ \bibinfo {pages} {eadl1814} (\bibinfo {year} {2024})}\BibitemShut {NoStop}%
\bibitem [{\citenamefont {McCormick}\ \emph {et~al.}(2008)\citenamefont {McCormick}, \citenamefont {Marino}, \citenamefont {Boyer},\ and\ \citenamefont {Lett}}]{PhysRevA.78.043816}%
  \BibitemOpen
  \bibfield  {author} {\bibinfo {author} {\bibfnamefont {C.~F.}\ \bibnamefont {McCormick}}, \bibinfo {author} {\bibfnamefont {A.~M.}\ \bibnamefont {Marino}}, \bibinfo {author} {\bibfnamefont {V.}~\bibnamefont {Boyer}},\ and\ \bibinfo {author} {\bibfnamefont {P.~D.}\ \bibnamefont {Lett}},\ }\bibfield  {title} {\bibinfo {title} {Strong low-frequency quantum correlations from a four-wave-mixing amplifier},\ }\href {https://doi.org/10.1103/PhysRevA.78.043816} {\bibfield  {journal} {\bibinfo  {journal} {Phys. Rev. A}\ }\textbf {\bibinfo {volume} {78}},\ \bibinfo {pages} {043816} (\bibinfo {year} {2008})}\BibitemShut {NoStop}%
\end{thebibliography}%
\end{document}